\documentclass[aps,prd,twocolumn]{revtex4}
\usepackage{amsmath,amssymb, amsthm}
\usepackage{graphicx,epsf}
\def\half{\frac{1}{2}}

\def\rf#1{(\ref{#1})}
\newcommand{\eqa}{\begin{eqnarray}}
\newcommand{\neqa}{\end{eqnarray}}
\newcommand{\bea}{\begin{eqnarray}}
\newcommand{\eea}{\end{eqnarray}}
\newcommand{\be}{\begin{equation}}
\newcommand{\ee}{\end{equation}}

\newcommand{\Ref}[1]{(\ref{#1})}

\renewcommand{\texttt}{{}}
\usepackage{graphicx}

\def\sdmass{\ensuremath{\sqrt{a_0}/2}}
\def\su2{\ensuremath{\mathfrak{su(2)}}}
\def\SU2{\ensuremath{{SU(2)}}}
\begin{document}

\title{%
%Self-Dual Loop Black Hole}
%Small-Large Distances dual Black Holes from Loop Gravity}
Self-dual Black Holes in LQG: \\ Theory \& Phenomenology}%Loop Gravity}
%Theory \& Phenomenology of Self-Dual Black Holes in LQG} %. Theory and Phenomenology}
%T-Dual Black Holes from Loop Gravity}%\\ \& \\
%Solution of Information Loss Problem}
%Mirror Black Holes
%Loop black hole space-time structure}%Space-time structure of the loop quantum black hole}
%\title{Space-time structure of the loop quantum black hole}%inside and outside the event horizon}

\author{Leonardo Modesto \& Isabeau Pr\'emont-Schwarz}

\affiliation{ Perimeter Institute for Theoretical Physics, 31 Caroline St.N., Waterloo, ON N2L 2Y5, Canada}

\date{\small\today}

\begin{abstract} \noindent
In this paper we have recalled the semiclassical metric
obtained from a classical analysis of the loop quantum black hole
(LQBH). We show that the regular Reissner-Nordstr\"om-like metric
is self-dual in the sense of T-duality: the form of the metric obtained
in Loop quantum Gravity (LQG) is invariant under the exchange $r \rightarrow a_0/r$
where $a_0$ is proportional to the minimum area in LQG and $r$ is the standard Schwarzschild radial coordinate at asymptotic infinity.
Of particular interest, the symmetry imposes that if an observer in $r\rightarrow +\infty$ sees a black hole of mass $m$
an observer in the other asymptotic infinity beyond the horizon (at $r \approx 0$) sees a dual mass
$m_P/m$. We then show that small LQBH are stable and could be a component of dark matter.
Ultra-light LQBHs created shortly after the Big Bang
would now have a mass of approximately $10^{-5} \, m_P$
%\times 10^{-3} m_P$
and emit radiation with a typical energy of about $10^{13} - 10^{14} {\rm eV}$ but they would also emit cosmic rays of much higher energies, albeit few of them. %Were
If these small LQBHs form a majority of the dark matter of the Milky Way's Halo,
the production rate of ultra-high-energy-cosmic-rays (UHECR) by these ultra light black holes would be compatible
with the observed rate of the Auger detector.
%would be $1.1\times 10^4$ ultra-high-energy cosmic rays $s^{-1} m^{-3}$; such abundance of ultra-high-energy cosmic rays not being observed, we deduce that LQBH do not make up the majority of dark matter.
%XXXXX
%LQBH created during the Big Bang would now have a mass of $7.6\times 10^{-3} m_P$ and emit radiation with a typical energy of $10^{23} eV$. Were these small LQBH to form a majority of the dark matter of the Milky Way's Halo, the rate of production of ultra-high-energy cosmic rays would be $1.1\times 10^4$ ultra-high-energy cosmic rays $s^{-1} m^{-3}$; such abundance of ultra-high-energy cosmic rays not being observed, we deduce that LQBH do not make up the majority of dark matter.

\end{abstract}

\maketitle
%\newpage
%\thispagestyle{plain}
\tableofcontents

%\listoffigures
%
%\newpage

\section*{Introduction}

Quantum gravity is the theory  attempting to reconcile general relativity and quantum
mechanics. %is one of  major problem in theoretical physics today.
In general relativity the space-time is dynamical,
it is then not possible to study other interactions on a fixed background because
the background itself is a dynamical field.
 %One of more diffuse
 %Among the quantum gravity theories,
 ``Loop quantum gravity" (LQG) \cite{book} \cite{SpinFoams} \cite{Fracton}
is a major contestant amongst the theories aiming at unifying gravity and quantum mechanics.
It is one of the non perturbative and
background independent approaches to quantum gravity.
%(an\includegraphics[]{../../Desktop/Entropy.eps}
%other non perturbative approach to quantum gravity is called  ``asymptotic safety
%quantum gravity" \cite{MR}).
%In this paper we will apply ideas suggested by full loop quantum gravity
%to a minisuperspace model where we will impose symmetries on the full
%metric to obtain a reduced model. It is possible to implement
%the Dirac quantization program following the fundamental ideas
%of loop quantum gravity.
%Loop quantum gravity
Since LQG is a quantum geometric
fundamental theory that reconciles general relativity
and quantum mechanics at the Planck scale, we expect that this theory could resolve the
classical singularity problems of general relativity.
Much progress has been done in this direction in the
last years.
In particular, the application of LQG technology
to the early universe in the context of minisuperspace models
have resolved the initial singularity problem \cite{Boj}, \cite{MAT}.

Black holes %interesting places where to test the validity of quantum gravity.
are another interesting place for testing the validity of LQG.
In the past years
applications of LQG ideas to the Kantowski-Sachs space-time
\cite{KS} \cite{h}
lead to some interesting results. % in this field.
In particular, it has been
shown %in black hole physics
\cite{work1} \cite{work2} that it is
possible to solve the black hole singularity problem by using
tools and ideas developed in the full LQG. %and applied to a minisuperspace model.
Other remarkable results have been obtained in the non homogeneous
case \cite{GP}. We think the resolution of the black hole singularity problem is
a crucial first step to solve the information loss problem \cite{Sabine}.

There is also work of a semiclassical nature which tries to solve the black hole singularity
problem \cite{SS},\cite{SS2}. %, \cite{SS2}.
 In these papers the authors use an effective Hamiltonian constraint
 obtained by replacing the Ashtekar connection $A$ with the holonomy $h(A)$
 and they solve the classical Hamilton equations of motion exactly
 or numerically.
 In this paper we try to improve on the semiclassical analysis by
 introducing a very simple modification to the holonomic version of the
 Hamiltonian constraint.
% from a
 %conservative point of view.
% In this paper we consider a conservative semiclassical analisys of LQBH
 %investigating the space-time structure.
 The main result is that
 the minimum area \cite{LoopOld} of full LQG is the
 fundamental ingredient to solve
 the black hole space-time singularity problem at $r=0$.
 The $S^2$ sphere bounces on the minimum area $8 \pi a_0$ of LQG
 and the singularity disappears.
 We show that the Kretschmann  invariant is regular in all of space-time and
 the position of the maximum is independent of the mass and of the polymeric parameter
 introduced to define the holonomic version of the scalar constraint.
 The radial position of the curvature maximum depends only on $G_N$ and $\hbar$.

This paper is organised as follows.  In the first section we
recall the singularity free semiclassical black hole solution obtained in \cite{RNR}.
%In the second section we recall the thermodynamics and we analyse
%the information loss problem
We also recall %using
the causal space-time structure
%structure supported by
and the Carter-Penrose diagram for the maximal space-time extension.
In the second section we show the self-duality property of the metric.
We take special notice of ultra-light black holes which differ qualitatively from Schwarzschild black holes even outside the horizon. We show that their horizons are hidden behind a wormhole of Planck diameter.
In the third section we study the phenomenology of LQBHs. We analyse the LQBH
%that they can be in a stable
thermodynamic and the relation with the cosmic microwave background.
We study the production rate of black holes in the early universe
%We suppose an initial value for the black hole mass created in the early universe
and using Stefan's law we calculate the black hole mass today.
We assume that the majority of dark matter is formed by ultra-light LQBHs
and consequently we estimate
the production of ultra-high-energy-cosmic-rays (UHECR). We show the production
of UHECR is compatible with observation. The ultra light black holes could be the missing source for the UHECRs.
% for the particular initial value of the black hole mass
%we have chosen.
%We conclude the section studying the black hole production in the early universe.
%Our data fit
%
%XXXXX
% and thus that they could possibly be the long sought after dark matter. However we show that given the currently estimated time of the universe, they would not have enough time to cool and would still be hot enough to produce an amount of ultra-high-energy cosmic rays much in excess of what we observe. This forces us to conclude that ultra-light black holes cannot form the majority of dark matter. In the fourth section we come back on the relationship between Loop Quantum Gravity and the LQBH.
%similar to the Reissner-Nordstr\"om solution for a black hole with
%mass and charge.

%\footnote{jjjjj}

%inside the black hole
% \cite{SS} and we extend the solution outside the event horizon
 %showing the regularity of the curvature invariant $\forall \,  r\geqslant 0$.
%In the second section we calculate the Hawking temperature
%and the entropy in terms of the event horizon area. In the same section
%we study also the mass evaporation process
%discussing the new physics suggested by loop quantum gravity.

\section{A regular Black Hole %Reissner-Nordstr\"om-like
from LQG}
In this section we recall the classical Schwarzschild solution inside the event
horizon $r \leqslant 2m$ \cite{RNR}, \cite{work1} \cite{work2}.
Because we are inside the event horizon the radial coordinate is
time-like and the temporal coordinate is space-like.
For this reason the space-time for $r \leqslant 2m$ is
the homogeneous %but non isotropic
Kantowski-Sachs
space-time of spatial topology $\mathbb{R} \times S^2$.
The Ashtekar's variables \cite{variables} are
\begin{eqnarray}
&& A= \tilde{c} \tau_3 d x + \tilde{b} \tau_2 d \theta - \tilde{b} \tau_1 \sin \theta d \phi + \tau_3 \cos \theta d
\phi,
\nonumber \\
&&E = \tilde{p}_c \tau_3 \sin \theta \frac{\partial}{\partial x} + \tilde{p}_b \tau_2 \sin \theta
\frac{\partial}{\partial \theta} - \tilde{p}_b \tau_1 \frac{\partial}{\partial \phi}.
\label{contriad}
\end{eqnarray}
The component variables in the phase space have length dimension:
$[\tilde{c}]=L^{-1}$, $[\tilde{p}_c]=L^{2}$, $[\tilde{b}]=L^{0}$, $[\tilde{p}_b]=L$.
Using the general relation $E^a_i E^b_j \delta^{i j}= {\rm det}(q) q^{ab}$ ($q_{ab}$ is the metric on
the spatial section) we obtain
$q_{ab} = (\tilde{p}_b^2/|\tilde{p}_c|, |\tilde{p}_c|, |\tilde{p}_c| \sin^2 \theta)$.
In the Hamiltonian constraint and in the symplectic structure we
restrict integration over $x$ to a finite interval $L_0$ and we rescale
the variables as follows:
$b=\tilde{b}$, $c= L_0 \tilde{c}$, $p_b = L_0 \tilde{p}_b$, $p_c=\tilde{p}_c$.
The length dimensions of the new phase space variables are:
$[c]=L^{0}$, $[p_c]=L^{2}$, $[b]=L^{0}$, $[p_b]=L^2$.
From the symmetry reduced connection and density
triad we can read the component variables in the phase space:
$(b, p_b)$, $(c, p_c)$,
with Poisson algebra $\{c, p_c \} = 2 \gamma G_N$, $\{b, p_b \} = \gamma G_N$.
%The solutions of equations (\ref{Eq.1}) using the time parameter $t \equiv e^T$
%and redefining the integration constant $\equiv e^{T_0} = 2 m$
%(see the papers in \cite{work1} \cite{work2})
The Hamiltonian constraint in terms of the rescaled phase space
variables and the holonomies is
\begin{eqnarray}
\hspace{0.05cm}\mathcal{C}_{H} \hspace{-0.05cm} = \hspace{-0.05cm} -
  \frac{N}{\kappa }
\Bigg\{  2 \frac{\sin \delta_c c}{\delta_c} \frac{\sin \Delta_b b}{\delta_b}  \sqrt{|p_c|}
+ \frac{\sin^2 \Delta_b b+ \gamma^2 \delta_b^2}{\sqrt{|p_c|} \delta_b^2} p_b \Bigg\},
\nonumber %\label{CH}
\end{eqnarray}
where $\kappa = 2 G_N \gamma^2$;
$\delta_b, \delta_c$ are
the polymeric parameters introduced to define
the lengths of the paths along which we integrate
the connection to define %the field in terms of
the holonomies
and by definition $\Delta_b=\delta_b/\sqrt{1+\gamma^2 \delta_b^2}$ \cite{RNR}.
The Gauss-constraint and the Diff-constraints are identically zero because of
the homogeneity. Using the gauge
$N = (\gamma \sqrt{|p_c|} \mbox{sgn}(p_c) \delta_b)/(\sin \Delta_b  b)$,
we can solve the Hamilton equation of motion and the the Hamiltonian constraint (see \cite{RNR} for
details):
${\mathcal C}_H(q_i) = 0$,
$ \dot{q}_i = \{q_i, C_H\}$;
where $q_i = (c, p_c, b, p_b)$
obtaining  a solution on the four dimensional phase space:
$(c(t), p_c(t), b(t), p_b(t))$.
The relations between the Ashtekar and metric variables is explicit in the following
line element: %$h_{ab} = \mbox{diag}(p_b^2/|p_c| L_0^2, |p_c|, |p_c|  \sin^2 \theta)$ ($m$ contains the gravitational constant parameter $G_N$).
%The line element is
\begin{eqnarray}
ds^2=-N^2 \frac{dt^2}{t^2} + \frac{p_b^2}{|p_c| \, L_0^2} dx^2 + |p_c| (\sin^2 \theta d \phi^2+d\theta^2).
\label{line-element}
\end{eqnarray}
%The solution of the system (\ref{HE}) is given in \cite{RNR} with all the details.
%here we recall the solution in the Schwarzschild coordinates.
In \cite{RNR} we have
calculated the solution inside the event horizon ($r<2m$) and because of the regularity of the
solution $\forall \, r$ we have analytically extended the solution to $0 < r < +\infty$.
In particular the Kretschmann invariant
($K = {\rm R}_{\mu\nu\rho\sigma} {\rm R}^{\mu\nu\rho\sigma}$)
is regular $\forall r$ and
it is possible to extend analytically the solution beyond the horizons
(because as will be recalled below, the new metric has an inside event horizon).
In \cite{RNR} we found regular coordinates in any patch where the metric
has a coordinate singularity showing explicitly that the metric is
regular everywhere and can be extended to all of space-time.

Because of the regularity of the metric, we can use the usual
Schwarzschild coordinates where $r$ is space-like and $t$ is time-like outside
the event horizon. The semiclassical metric is
%We can write the metric in another form which is more %evident the
%similar to the Reissner-Nordstr\"om space-time.
%The metric can be written in the following form
\begin{eqnarray}
&& \hspace{-0.5cm} ds^2 = -\frac{ (r - r_+) (r-r_-)(r+ r_{\star})^2 }{r^4 + a_0^2}dt^2 \nonumber \\
&&\hspace{0.5cm}+\frac{dr^2}{\frac{(r-r_+)(r-r_-)r^4}{(r+ r_{\star})^2 (r^4 + a_0^2)}} +\Big(\frac{a_0^2}{r^2} +
r^2\Big) d\Omega^{(2)},
\label{metricabella}
\end{eqnarray}
where
$r_+=2m$, $r_- = 2m {\mathcal P}(\delta_b)^2$, $r_{\star} = 2m {\mathcal P}(\delta_b)$,
$a_0 = A_{\rm Min}/8 \pi$ and $A_{\rm Min}$ is the minimum area of LQG. ${\mathcal P}(\delta_b)$
is a function of the polymeric parameter $\delta_b$,
\begin{eqnarray}
{\mathcal P}(\delta_b) =\frac{\sqrt{1+\gamma^2 \delta_b^2} -1}{\sqrt{1+\gamma^2 \delta_b^2} +1}.
\end{eqnarray}
The area operator in LQG has a discrete spectrum, irreducible units of area --- associated to an edge on a spin-network ---  in LQG have area $A(j):= 8\pi\gamma \sqrt{j(j+1)} l_P^2$ where $\gamma$ is the Immirzi parameter believed to be $\gamma=0.2375$ \cite{gamma}, $j$ is a half-integer labelling an irreducible representation of \SU2 and $l_P$ is the Planck length. Looking at this, it is natural to assume that the minimum area in LQG is $A_{min}=A(1/2)= 4\pi\gamma \sqrt{3} l_P^2\approx 5 l_P^2$. One should however not take this exact value too seriously for various reasons. To mention but a few reasons, we have that: for one the value of $\gamma$ is not necessarily definite and the consensus on its value has change a few times already; second there are other Casimirs possible than $\sqrt{j(j+1)}$;   third, we are looking for a minimum area for a closed surface so, the spin-network being most likely a closed graph, it is probable that least two edges cross the surface, in which case the minimum area is at least twice the previously given value, in addition, if we consider a surface inclosing a non-zero volume, LQG stipulates that at least one 4-valent vertex must be present, in which case we might have for edges intersecting the surface making $A_{min}$ be four times the aforementioned value. We will parameterise our ignorance with a parameter $\beta$ and define
$A_{min}= \beta A(1/2)= 4\pi\gamma\beta \sqrt{3} \, l_P^2\approx 5\beta l_P^2$, and so $a_0=A_{min}/(8\pi)= \gamma\beta \sqrt{3}\, l_P^2/2\approx 0.2\beta l_P^2$ where the expectation is that $\beta$ is not many orders of magnitude bigger or smaller than $1$, in this article we mostly consider $\beta\approx 1$ or $\beta=4$ when more precision is need, but in the end the precise choice of $\beta$ does not change much.

\begin{figure}
 \begin{center}
 \hspace{-0.4cm}
  \includegraphics[height=4.5cm]{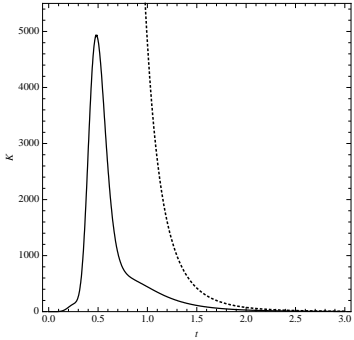}
    \end{center}
  \caption{\label{K0}
 Plot of the
 Kretschmann  scalar
  invariant ${\rm R}_{\mu \nu \rho \sigma} {\rm R}^{\mu \nu \rho \sigma}$
for $m = 10$, $p_{b}^{0} =1/10$ and $\gamma \delta_b = \log(4)/\pi$, $\forall t \geqslant 0$; the large
 $t$ behaviour %of the invariant
  is $1/t^6$.}% as shown in the zoom on the right side.}
  \end{figure}

There is also {\em another argument} we can make to justify the analytical extension of the metric to all of
space-time. %Using the solution (\ref{metricabella})
We can interpret our black hole solution (\ref{metricabella})
has having been generated by an effective
matter fluid that simulates the loop quantum gravity corrections (in analogy with \cite{BR}).
The effective gravity-matter system satisfies by definition the Einstein
equations $G =8 \pi \,T$, where $T$ is the
effective energy tensor. In this case $T \neq 0$ contrarily to the classical Schwarzschild solution.
The stress energy tensor for a perfect fluid compatible with the space-time
symmetries is $T^{\mu}_{\nu} = (- \rho, P_r, P_{\theta}, P_{\theta})$
and in terms of the Einstein tensor the components are
$\rho= - G^t_t/8 \pi G_N$, $P_r = G^r_r/8 \pi G_N$
and $P_{\theta}= G^{\theta}_{\theta}/8 \pi G_N$.
The semiclassical metric to zeroth order in $\delta_b$ and $a_0$
is the classical Schwarzschild solution ($g_{\mu \nu}^{C}$)
that satisfies $G^{\mu}_{\nu}(g^C) \equiv 0$.
When we calculate explicitly the energy density and pressure
we obtain that those quantities are {\em spatially homogeneous} inside the
event horizon and {\em static} outside.
Using this property of the energy tensor we can repeat the
argument used to extend the classical Schwarzschild solution to all of space-time.
The crucial difference is that in our case $T^{\mu}_{\nu} \neq 0$ but the logic is identical.
In particular  $T^{\mu}_{\nu}$ is {\em static} or {\em spatially homogeneous} %timelike or spacelike %or null
depending on the nature of the surfaces
$\sqrt{|p_c|} = const.$ and we can repeat the analysis given at the end of \cite{HE}.
The analytical form of the energy density is,
\begin{eqnarray}
&&\rho = 4 r^4 [a_0^4 m (1 + {\mathcal P})^2 +
   2 m^2 {\mathcal P} (1 + {\mathcal P})^2  r^7 +\nonumber \\
&& \hspace{0cm} - a_0^2  r^2 (2 m {\mathcal P} + r) (12 m^2 {\mathcal P}^2 -
      m (7 + {\mathcal P} (2 + 7 {\mathcal P})) r \nonumber \\
      &&+ 3 r^2)]/[8 \pi G_N (2 m {\mathcal P} + r)^3 (a_0^2 +
    r^4)^3].
\label{edensity}
\end{eqnarray}
\begin{figure}
 \begin{center}
  \includegraphics[height=5cm]{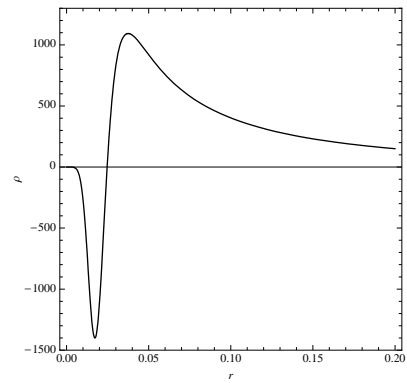}
  \hspace{1cm}
  \end{center}
  \caption{\label{energy} Effective energy density for $m=10$ and $a_0=0.01$.}
  \end{figure}
  \begin{figure}
 \begin{center}
  \includegraphics[height=3.4cm]{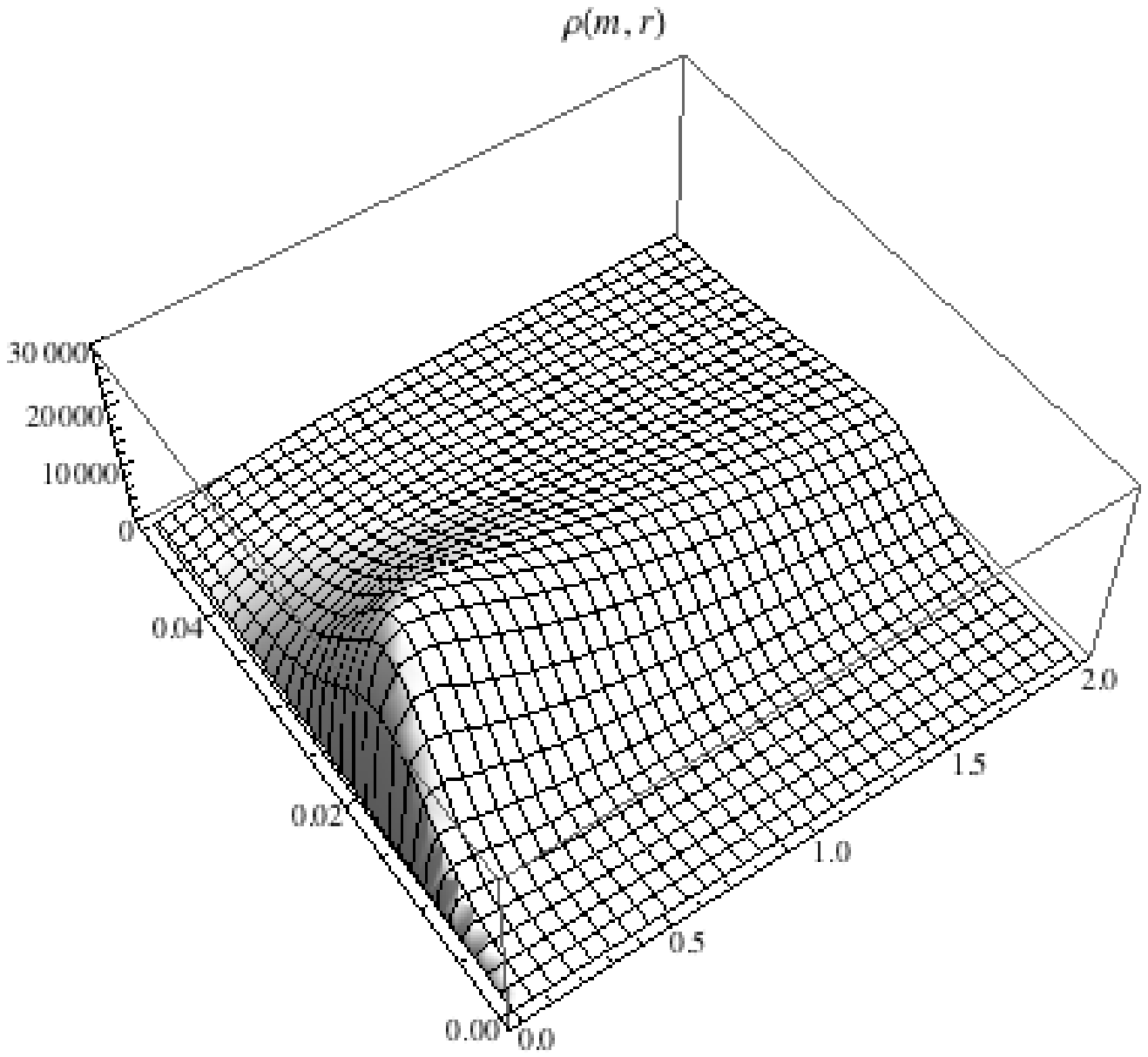}
    \includegraphics[height=3.2cm]{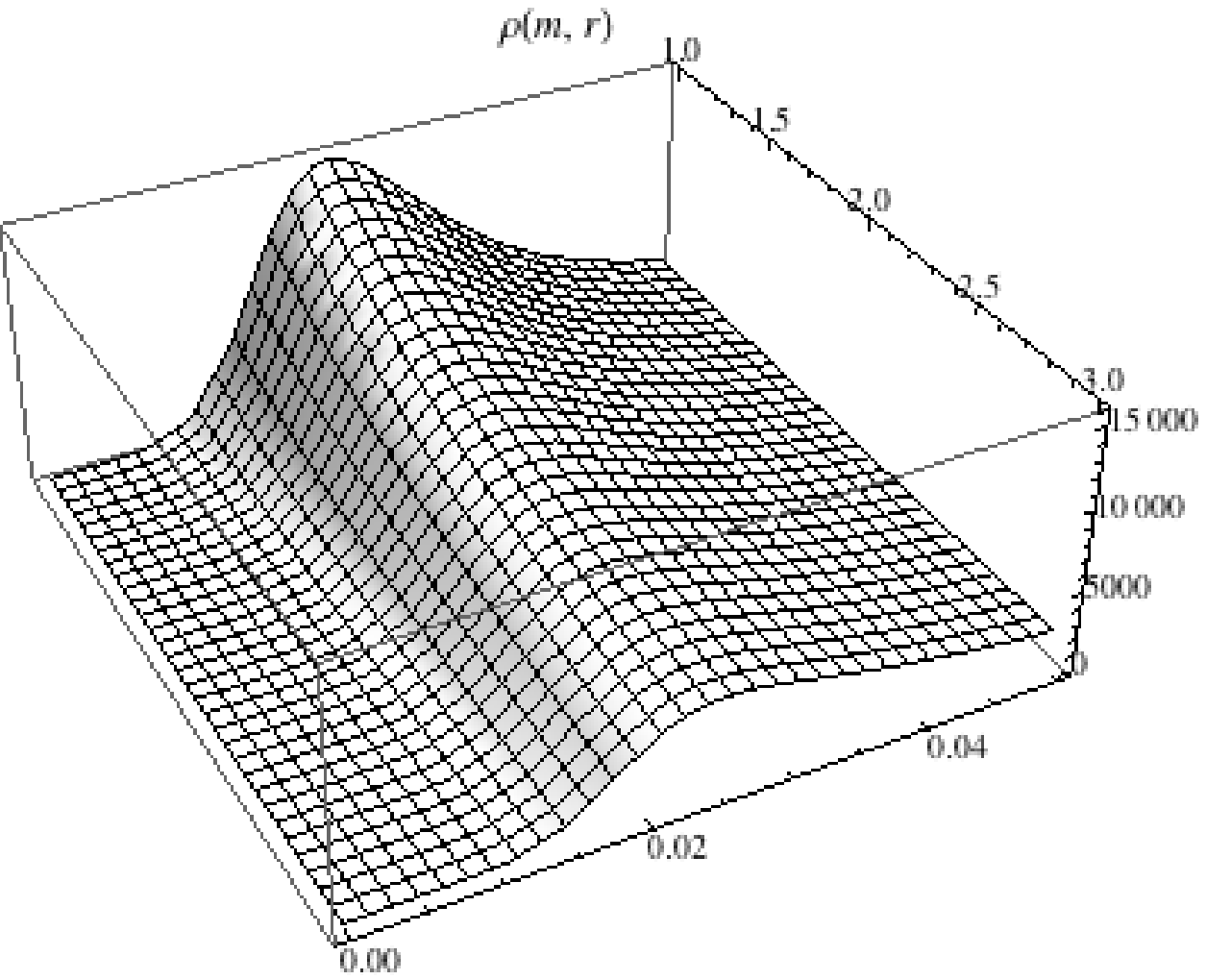}
      \includegraphics[height=3.4cm]{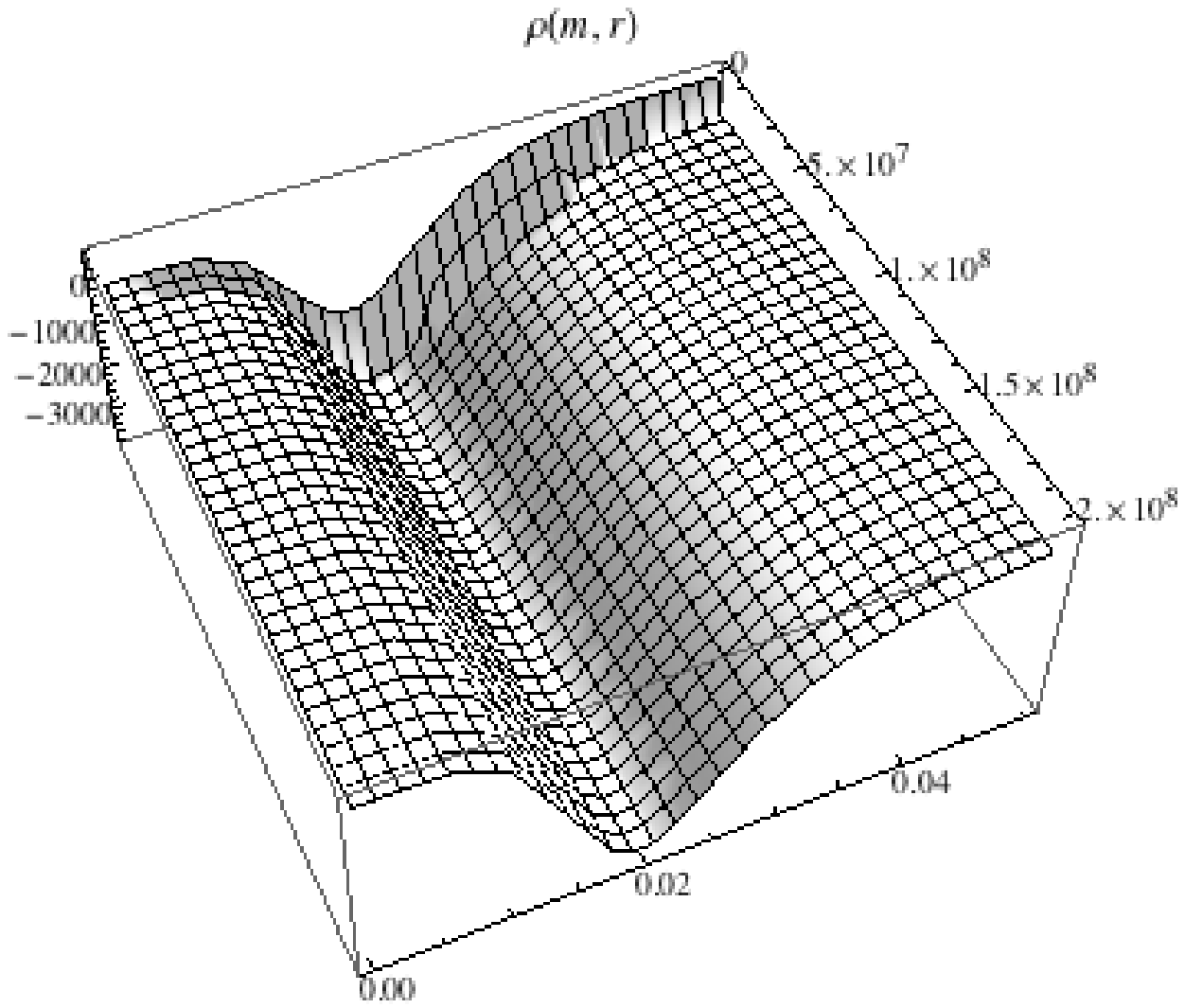}
  \hspace{1cm}
  \end{center}
  \caption{\label{energyrm}
  Effective energy density as function of $r$ end $m$. In the upper plot on the left
  $m\in[0, 2]$ and $r\in[0, 0.045]$, in the upper plot on the right $m\in[1, 3]$ and $r\in[0, 0.045]$
  and in the lower plot $m\in[0, 2 \, 10^8]$ and $r\in[0, 0.045]$. The plots show that
  the energy density is localised around the Planck scale for any value of the mass
  and decrees rapidly for $r\gtrsim l_P$.}
  \end{figure}
The regular properties of the metric are summarized
in the following table,
\begin{center}
\begin{tabular}{|r|}
\hline
$ \,\,\,\,\, {\rm Properties \,\, of} \,\, g_{\mu \nu}$ \,\,\,\, \\
\hline
\hline
$\lim_{r \rightarrow +\infty} g_{\mu \nu}(r) = \eta_{\mu \nu}$  \\
\hline
\hline
$\lim_{r \rightarrow 0} g_{\mu \nu}(r) = \eta_{\mu \nu}$   \\
\hline
\hline
$\lim_{m, a_0 \rightarrow 0} g_{\mu \nu}(r)  = \eta_{\mu \nu}$ \\
\hline
\hline
$K(g) < \infty \,\, \forall r$\\
\hline
\hline
$r_{\rm Max}(K(g)) \propto l_P$ \\
\hline
\end{tabular}
\label{good}
\end{center}
Where $r_{\rm Max}(K(g))$ is the radial position of the where the Kretschmann invariant achieves its
maximum value. Fig.\ref{K0} is a graph of $K$, it is regular in all of space-time and the large
$r$ behaviour is asymptotically identical to the classical singular scalar
${\rm R}_{\mu \nu \rho \sigma} {\rm R}^{\mu \nu \rho \sigma} = 48 m^2/r^6$.
The resolution of the regularity of $K$ is a non perturbative result, in fact for small values of the
 radial coordinate $r$,
$K\approx 3145728 \pi^4 r^6/a_0^4 \gamma^8 \delta_b^8 m^2$ diverges for $a_0\rightarrow 0$
or $\delta_b \rightarrow 0$.
A crucial difference with the classical Schwarzschild solution is that
the the 2-sphere $S^2$ has a minimum for
$r_{min} = \sqrt{a_0}$
and the minimum square radius is $p_{c}(r_{min}) = 2 a_0$.
The solution has a spacetime structure very similar to the
Reissner-Nordstr\"om metric because of the inner horizon in
$r_{-} = 2 m {\mathcal P}(\delta_b)^2$.
 For $\delta_b \rightarrow 0$, $r_{-} \approx m \gamma^4 \delta_b^4/8$.
We observe that the position of the inside horizon is $r_-\neq 2m \,\, \forall \gamma \in\mathbb{R}$
(we recall that $\gamma$ is the Barbero-Immirzi parameter).
The metric (\ref{metricabella}) for
$\delta_b, a_0 = 0$ is exactly the Schwarzschild metric.

The metric (\ref{metricabella}) has an asymptotic Schwarzschild
core near $ r \approx 0$. To show this property
we develop the metric very close to the point $r\approx 0$ and we consider the
coordinate changing  $R= a_0/ r$.
In the new coordinate the point $r=0$ is mapped
in the point $R=+ \infty$.
The metric in the new coordinates is
\begin{eqnarray}
ds^2= - \bigg(1- \frac{2 m_1}{R} \bigg) dt^2
+ \frac{dR^2}{1- \frac{2 m_2}{R}} + R^2 d \Omega^{(2)},
\label{metricar0b}
\end{eqnarray}
where $m_1$ and $m_2$ are functions of $m, a_0, \delta_b, \gamma$,
\begin{eqnarray}
 m_1=  \frac{a_0 }{8 \pi m \gamma^2 \delta_b^2 {\mathcal P}(\delta_b)}\,\, , \,\,\,\,
 m_2=  \frac{a_0  (1 + \gamma^2 \delta_b^2)}{8 \pi m \gamma^2 \delta_b^2 {\mathcal P}(\delta_b)}.
 \label{m1m2}
\end{eqnarray}
For small $\delta_b$ we obtain $m_1\approx m_2$ and (\ref{metricar0b}) converges to
a Schwarzschild metric of mass
$M \approx a_0/ 2 m \pi \gamma^4 \delta_b^4$. We can conclude the space-time
near the point $r\approx 0$ is described by an effective Schwarzschild metric
of mass $M\propto a_0/m$
in the large distance limit $R\gg M$.
%with mass $M\sim a_0/m$.
%and radial coordinate $R$ in the limit $R\ggg M$.
An observer in the asymptotic region $r=0$ experiences a Schwarzschild metric
of mass $M\propto a_0/m$.
  %%%%%%%%%%%%%%%%%   6)

\begin{figure}
 \begin{center}
  \includegraphics[height=3cm]{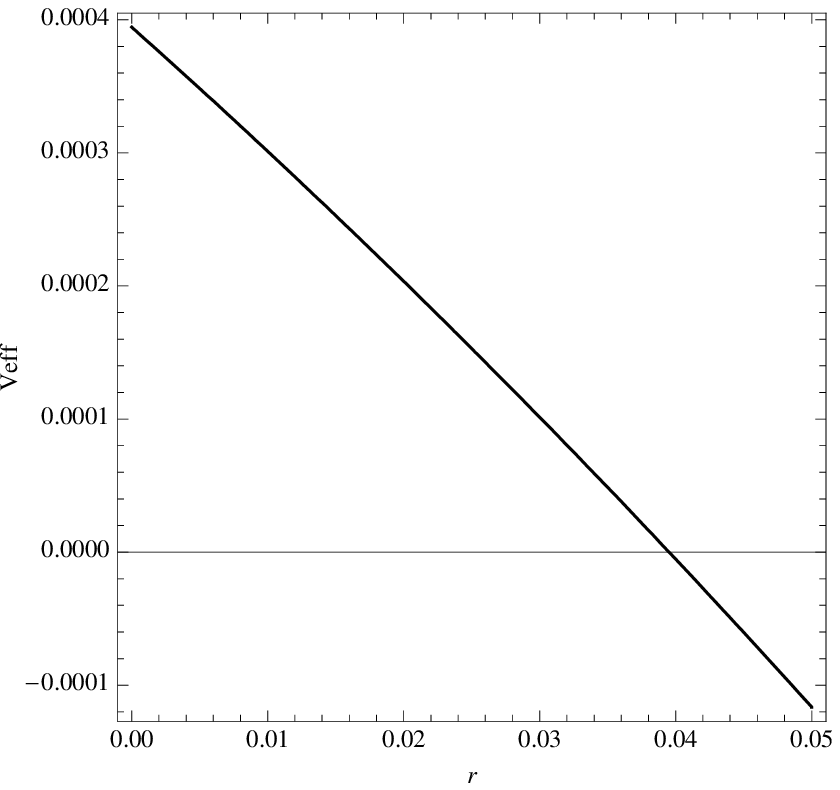}
   \includegraphics[height=3cm]{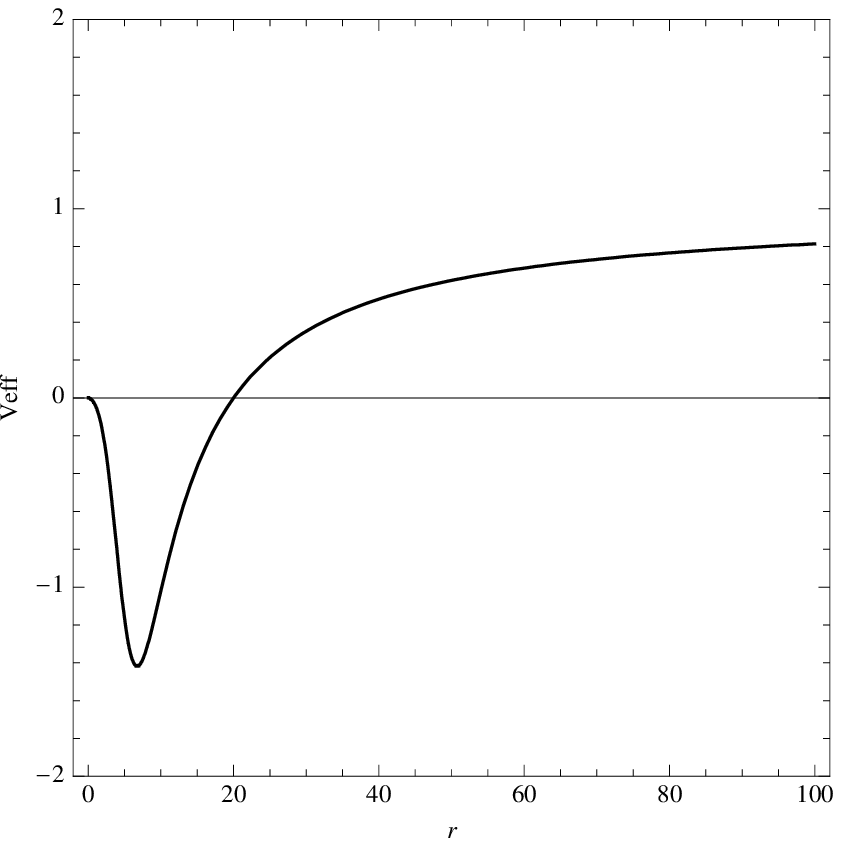}
  %\hspace{1cm}
  \end{center}
  \caption{\label{Veff}
  Plot of $V_{eff}(r)$. On the left there is a zoom of $V_{eff}$ for
  $r\approx 0$.}
  \end{figure}

\begin{figure}
 \begin{center}
  \includegraphics[height=2.5cm]{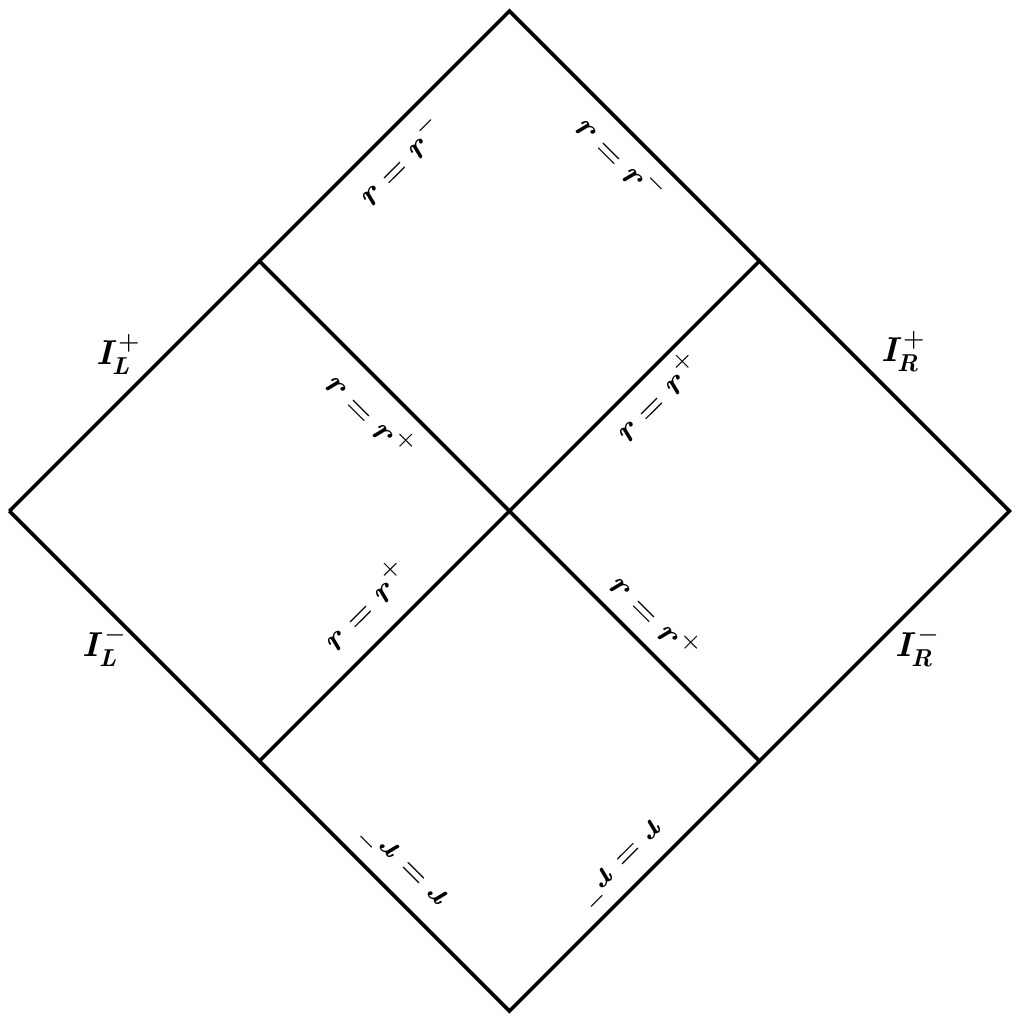}
  \hspace{0.7cm}
  \includegraphics[height=2.5cm]{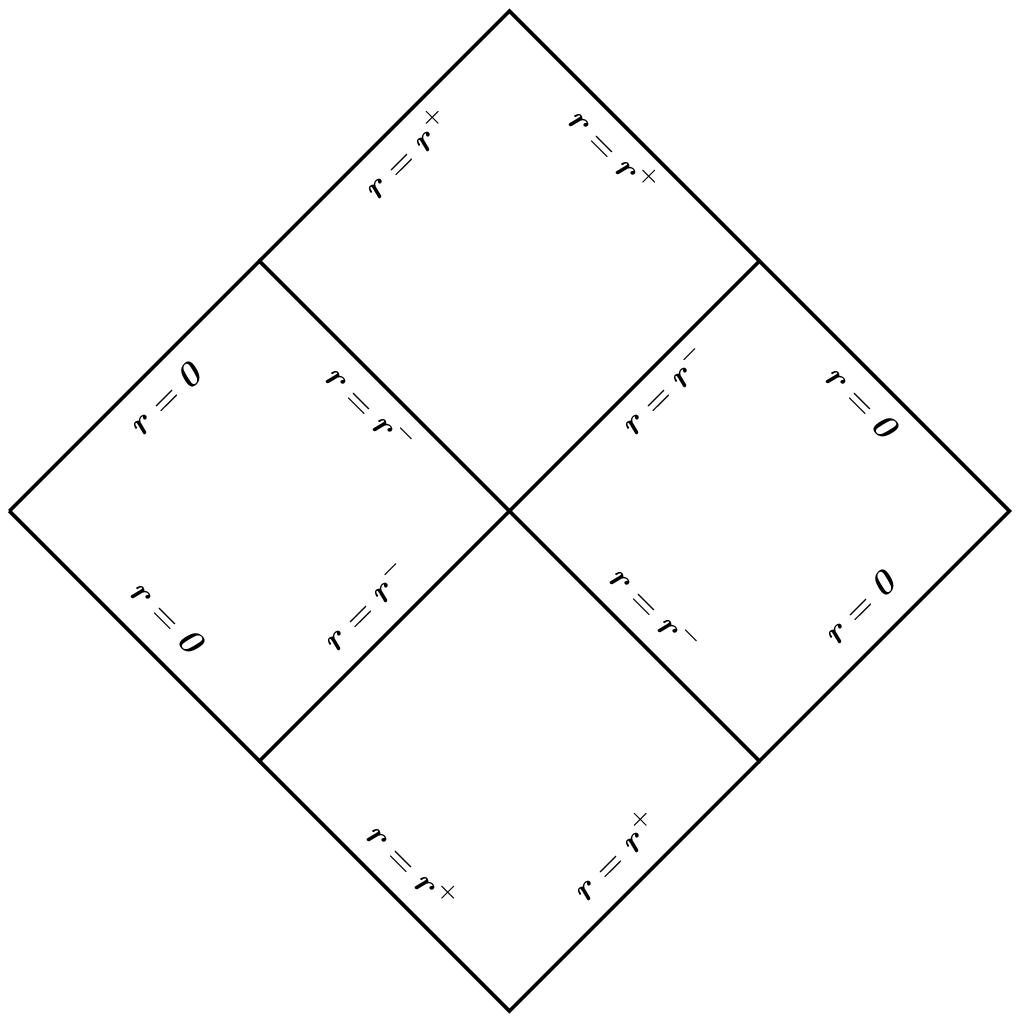}
  %\vspace{1cm}
  \includegraphics[height=6cm]{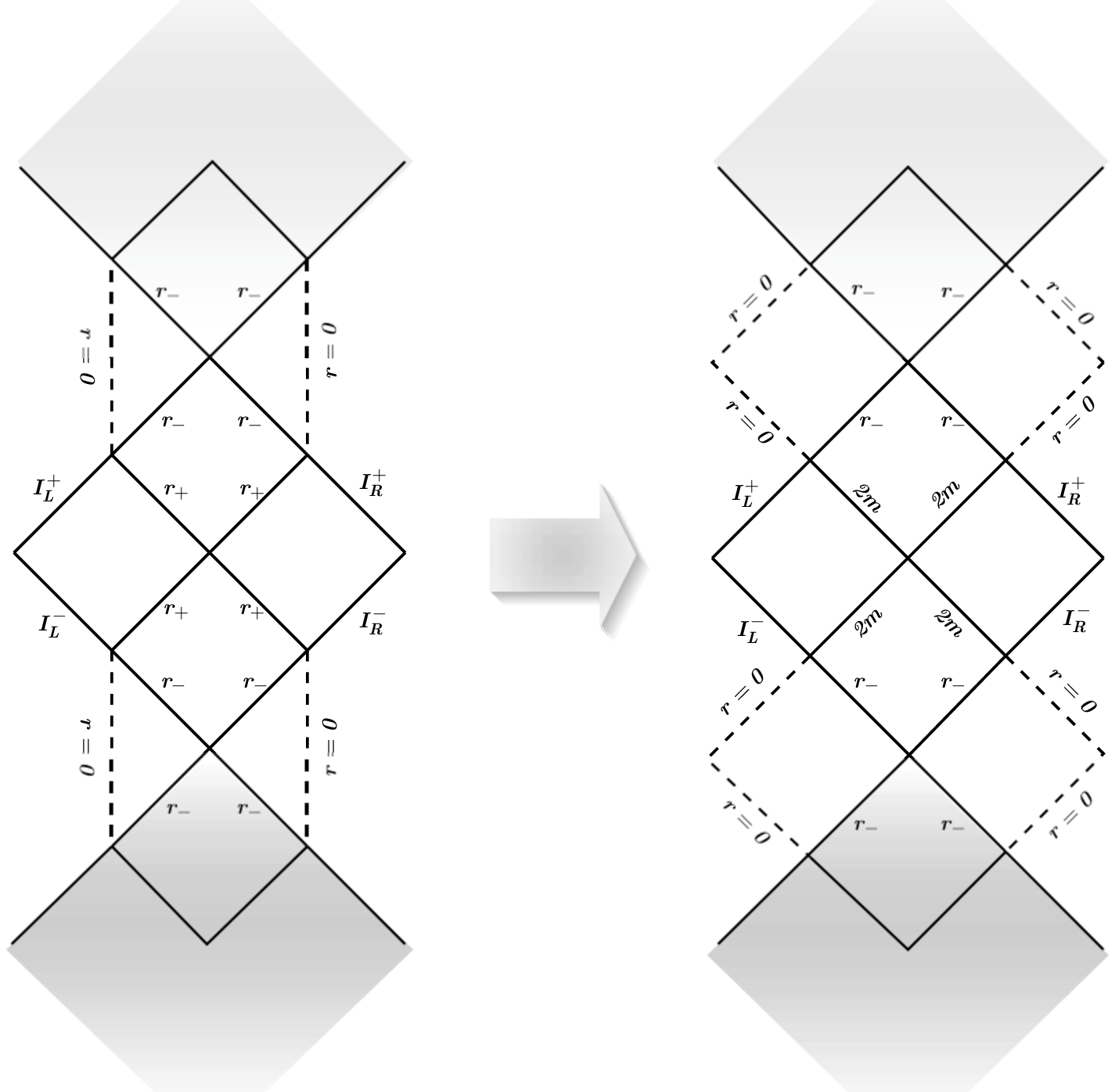}
  \hspace{1cm}
  \end{center}
  \vspace{-0.5cm}
  \caption{\label{penrose}
  The upper picture on the left represents the Carter-Penrose diagram in
  the region outside $r_-$ and the upper picture on the right represents the diagram
  for $r_-\leqslant r \leqslant 0$.
  The lower pictures represent the maximal space-time extension of the LQBH on the right and the analog extension
  for the Reissner-Nordstr\"om black hole.}
%  \vspace{-1cm}
  \end{figure}

  Now we are going to show that a massive particle can not reach
  %The region
  $r=0$ in a finite proper time.
  %could not be in a finite proper time as we go to show.
  We consider the radial geodesic equation for a massive point particle
  \begin{eqnarray}
 (- g_{tt} \, g_{rr}) \dot{r}^2 = E_n^2 + g_{tt},
  \label{geometricabella}
  \end{eqnarray}
  where  ``$\,\, \dot{} \,\,$"  is the proper time derivative and $E_n$ is the point particle
  energy. If the particle falls from infinity with zero initial radial velocity
  the energy is $E_n=1$.
  We can write (\ref{geometricabella}) in a more familiar form
    \begin{eqnarray}
   \underbrace{(- g_{tt} \, g_{rr})}_{\geqslant 0 \,\, \forall r} \dot{r}^2 + \underbrace{V_{eff}}_{-g_{tt}}(r) =
   \underbrace{E}_{E_n^2},
  \label{geometricabella2}
  \end{eqnarray}
$V_{eff}$ is plotted in Fig.\ref{Veff}.
For $r=0$, $V_{eff}(r=0) = 4 m^4 \pi^2 \gamma^8 \delta_b^8/a_0^2$
(for small $\delta_b$)
then a particle with $E<V_{eff}(0)$ can never reach $r=0$.
If the particle energy is $E_n>V_{eff}(0)$, the geodesic equation
for $r\approx 0$ is
$\dot{r}^2 \approx r^4$ which gives $\tau \approx 1/r - 1/r_0$ or
$\Delta \tau \equiv \tau(0) - \tau(r_0) \rightarrow +  \infty$ for the proper time to reach $r=0$.
  The space-time structure of the semiclassical solution is given in Fig.\ref{penrose}.
%We can compose the diagrams in Fig.(\ref{outrmeno}) to obtain a maximal extension similar
  %to the Reissner-Nordstr\"om one, the result is represented in Fig.(\ref{penrose}).
  \section{Selfduality}% and The Dual World}
 In this section we explicitly show that the black hole solution obtained in LQG is {\em selfdual}
 in the sense the metric is invariant under the transformation
 $r\rightarrow a_0/r$.
  The self-dual transformation will transform the relevant quantities as shown in the following table:
 \begin{center}
\begin{tabular}{|r|}
\hline
$ {\rm Self-duality} $  \\
\hline
 $r \rightarrow R= \frac{a_0}{r}$ \\
\hline
$  r_+ \rightarrow R_- = \frac{a_0}{r_+} = \frac{a_0}{2 m } $  \\
\hline
\hline
$r_- \rightarrow R_+ = \frac{a_0}{r_-}  = \frac{a_0}{2 m {\mathcal P}(\delta_b)^2}$   \\
\hline
\hline
$r_{\star} \rightarrow R_{\star} = \frac{a_0}{r_{\star}}
= \frac{a_0}{2 m {\mathcal P}(\delta_b)}$ \\
\hline
\hline
$m \rightarrow M = \frac{a_0}{m {\mathcal P}(\delta_b)^2}$ \\
\hline
%\hline
\end{tabular}
\label{sd}
\end{center}
(note that $R_+ > R_-$ $\forall \delta_b$ because ${\mathcal P}(\delta_b) <1$).
If we apply to this transformation to the metric (\ref{metricabella}), we obtain
\begin{eqnarray}
&& \hspace{-0.5cm} ds^2 = -\frac{ (R- R_+) (R-R_-)(R+ R_{\star})^2 }{R^4 + a_0^2}dt^2 \nonumber \\
&&\hspace{0.2cm}+\frac{dR^2}{\frac{(R-R_+)(R-R_-)R^4}{(R+ R_{\star}^d)^2 (R^4 + a_0^2)}} +\Big(\frac{a_0^2}{R^2} +
R^2\Big) d\Omega^{(2)},
\label{metricabellad}
\end{eqnarray}
where we have complemented the transofmation $r\rightarrow a_0/r$ with a rescaling of the time coordinate
$t \rightarrow {\mathcal P}(\delta_b) ({r_+^{3/2} r_-^{1/2}}/a_0) \, t$.
It is evident from the explicit form (\ref{metricabellad}) that the metric is {\em selfdual}.
% \subsection{The dual world}
 We can define the dual Schwarzschild radius %(or the dual mass $m^d$)
 identifying $R=a_0/r$, $r^{sd}=\sqrt{a_0}$. The existence of a selfdual radius
 implies a selfdual mass because we have
 \begin{eqnarray}
\hspace{-0.2cm} R_- = r_- \,\, , \,\, R_+=r_+ \,\, , \,\, R_{\star} = r_{\star} \rightarrow
 m_{sd} = \frac{\sqrt{a_0}}{2{\mathcal P}(\delta_b)}.
\label{dualm}
 \end{eqnarray}
 % The duality relation between the outer event horizon radius of the black hole and the
% event horizon of the dual black hole is: $r_-^d r_+= a_0/{\mathcal P}(\delta_b)^2$.
% The selfdual radius is $r^{sd}=\sqrt{a_0}/{\mathcal P}(\delta_b)^2$ and the selfdual mass
%can be obtained identifying $m = m^d$, $m^{sd} = \sqrt{a_0}/2{\mathcal P}(\delta_b)$.
In the global extension of the space-time any black hole with mass
$m < m_{sd}$ is equivalent to a black hole of mass $m > m_{sd}$ by the selfdual symmetry.
%At this point we want to introduce the two points function for a scalar field in
%our fixed space-time.
%In spherical coordinate and using the dual symmetry the Green function is
%\begin{eqnarray}
%\langle \phi(x) \phi(0) \rangle \sim \frac{1}{\frac{a_0^2}{x^2} + x^2 + \alpha (\frac{a_0}{x} + x ) + \beta}
%\label{2p}
%\end{eqnarray}
%%%%%%%%%%%%%%%%%%%%%%%%%%%%%%%%%%%%%%%%%%%%%%%%%%%%%%%%
\subsection{Ultra-light LQBHs}%G Black Holes and Dark Matter}
\begin{figure}[tbp]
\begin{equation*}
\leavevmode\hbox{\epsfxsize=8.5 cm
   \epsffile{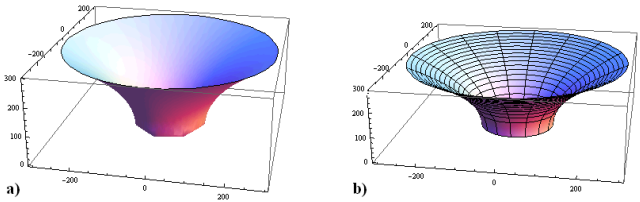}}
\end{equation*}%
\caption{Embedding diagram of a spatial slice just outside the horizon of a $50$ Planck mass ($\approx 1 {\rm mg}$) black hole.
In (a) we have the LQBH with metric (\ref{metricabella}); in (b) is the Schwarzschild black hole. In both
cases the foliation is done with respect to the time-like Killing vector and the scales are in Planck units. The
lowermost points in each diagram correspond to the horizon (the outer horizon in the LQG case).}
\label{biglqgbh}
\end{figure}

%If black holes are deemed to have the Schwarzschild metric, it can be argued that due to the quantum nature of particles, they do not form black holes. More precisely, it is argued that the de Broglie wave lengths of particles are greater than their Schwarzschild radius and thus, not enough mass is contained inside the Schwarzschild radius to form a black hole. The de Broglie wave length of a particle is $\lambda_{B}$ This argument is not as straightforward in the case of the LQG black hole.
Outside the exterior horizon, the LQBH %G black hole
metric (\ref{metricabella}) differs from the Schwarzschild metric only
by Planck size corrections. As such, the exterior of heavy LQBHs (where by ``heavy" we mean significantly
heavier than the Plank mass which is of the order of $20  \, \mu {\rm g}$) is not qualitatively different than that of a
Schwarzschild black hole of the same mass. This is shown in Fig.\ref{biglqgbh} where the embedding diagrams of the
LQG and Schwarzschild black holes of $50$ Planck masses are compared just outside the horizon.

\begin{figure}[tbp]
\begin{equation*}
\leavevmode\hbox{\epsfxsize=8.5 cm
   \epsffile{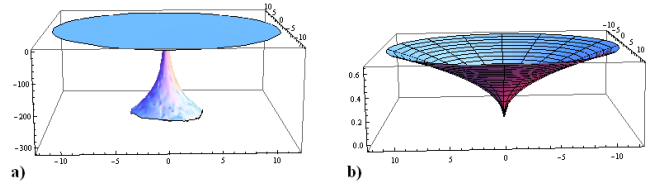}}
\end{equation*}%
\caption{Embedding diagram of a spatial slice just outside the horizon of a 0.005 Planck mass ($\approx 100 ng$) black
hole. In (a) we have the LQBH with metric (\ref{metricabella}); in (b) is the Schwarzschild black hole. In
both cases the foliation is done with respect to the time-like Killing vector and the scales are in Planck units. The
lowermost points in each diagram correspond to the horizon (the outer horizon in the LQG case).}
\label{smalllqgbh}
\end{figure}

\begin{figure}[tbp]
\begin{equation*}
\leavevmode\hbox{\epsfxsize=8.5 cm
   \epsffile{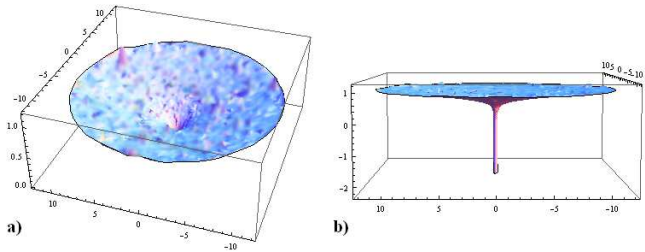}}
\end{equation*}%
\caption{(a) Embedding diagram of a spatial slice just outside the throat of a 0.005 Planck mass LQBH. (b)
zoom on the upper part of the throat of the same black hole. In both cases the foliation is done with respect to the
time-like Killing vector and the scales are in Planck units.}
\label{smalllqgbhzoom}
\end{figure}

In order to see a qualitative departure from the Schwarzschild black hole outside the horizon we must consider the
``sub-Planckian" regime, when the mass of the black hole is less than the Planck mass, as that is when quantum effects
will become significant. Due to Planck scale corrections the radius of the 2-sphere is not $r$, like is the case for
the Schwarzschild metric, but looking at (\ref{metricabella}) we see that the radius of the 2-sphere is
\begin{eqnarray}
\rho = \sqrt{r^2 + \frac{a_0^2}{r^2}} .
\label{rho}
\end{eqnarray}
We see that $\rho$ has a bounce at $r = \sqrt{a_0}$ which comes from LQG having a discrete area spectrum and thus a
minimal area (here $8 \pi a_0 $).  If the bounce happens before the outer horizon is reached, the outer horizon will be
hidden behind the Planck-sized wormhole created where the bounce takes place.  As a consequence of this, even if the
horizon is quite large (which it will be if $m<<m_P$) it will be invisible to observers who are at $r>\sqrt{a_0}$ and
who cannot probe the Planck scale because these observers would need to see the other side of the wormhole which has a
diameter of the order of the Planck length.  From this we conclude that to have this new phenomenon of hidden horizon
we must have $2m= r_+ < \sqrt{a_0}$, or $m<\sqrt{a_0}/2$. We illustrate this phenomenon with the embedding diagrams of
a LQBH of mass $m=4\pi \sqrt{a_0}/100$ in Fig.\ref{smalllqgbh}a
and Fig.\ref{smalllqgbhzoom} which
can be contrasted with the embedding diagram of the Schwarzschild black hole of the same mass in
Fig.\ref{smalllqgbh}b.

The formation of such ultra-light LQBHs is also of interest. For the Schwarzschild black hole, black hole formation occurs once a critical density is reached, i.e. a mass $m$ is localised inside a sphere of area $4\pi (2m)^2$. The ``heavy" LQBH is analogous: to create it we must achieve a critical density, putting a mass $m\geq \sqrt{a_0}/2$ inside a sphere of area $[4\pi (2m)^2 + a_0/(2m)^2]$. The requirement for the formation of an ultra-light LQBH is something else altogether because of the worm-hole behind which it hides: we must localise mass/energy (a particle or a few particles), irrespective of mass as long as the total mass satisfies $m< \sqrt{a_0}/2$) inside a sphere of area $8\pi a_0 $ as this ensures that the mass will be inside the mouth of the wormhole. Because $A_{min}\geq 5 l_P^2$ for any natural $\beta$ at the currently accepted value of the Immirzi parameter, there does not seem to be any semi-classical impediment to doing that. Hence it should be possible to create ultra-light black holes.
%This however is semiclassically impossible as the Compton wave length of a mass smaller than $\sqrt{a_0}/2$ is greater than $\sqrt{a_0}$. Though this is just a heuristic argument and the analysis of the actual collapse is needed to guarantee that an ultra-light LQBH can not be formed through collapse, this heuristic argument is very convincing because it's shakiest element is the use of the Compton wavelength in a curved space-time context
%\footnote{
%(the Compton wavelength relies on the commutation relationships of flat space-time as well as time translation invariance) but due to the light masses we are considering, the curvature is extremely small right up to the mouth of the wormhole. This does not mean however that ultra-light LQBHs can not exist, as they might still be formed as primordial black holes as well as from the evaporation of black holes formed by the collapse of a mass greater than $\sqrt{a_0}/2$.

 \subsection{``Particles-Black Holes" Duality}
 {\em Classical Duality.}
In this section we want to to emphasize the physical meaning of the duality
emerging from the self-dual metric analyzed in the paper.
The metric (\ref{metricabella})
describes a space-time with two asymptotic regions, the $r\rightarrow +\infty$ ($\equiv I^+$)
region and the $r\rightarrow 0$ ($\equiv I^0$) region. Two observers in the two regions
see some metric but they perceive two different masses. The observer in $I^+$
perceives a mass $m$, the observer in $I^0$ a mass $M\propto a_0/m$. Physically
any observed supermassive black hole in $I^+$ is perceived as a
an ultra-light ($m\ll m_P$) particle in $I^0$ and vice versa.
\begin{figure}
 \begin{center}
 \hspace{0.1cm}
  \includegraphics[height=3.5cm]{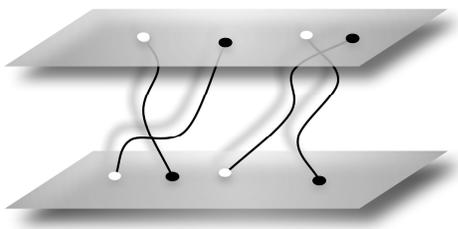}
      \end{center}
  \caption{\label{PBHD}
 Particle-Black Hole Duality in the Multiverse, an artistic picture.
  }
%\label{temperature}
  \end{figure}
The ultra-light particle is confined beyond the throat discussed in the previews
section because if $m\ll m_P$ then $r_+\ll \sqrt{a_0}$,
which is the throat radius or equivalently the self-dual radius.
This property of the metric leaves open the possibility to have a
{\em ``Quantum Particle-Black Hole" Duality}, in fact
a particle with $\lambda_c  \approx \hbar/2m \gg l_P$ could have sufficient space in $r <r_+$
because the physical quantity to compare with $\lambda_c$ is
$D = 2 [(2 G_N m)^2 + (a_0)^2/(2 G_N m)^2]^{1/2}$ %= \sqrt{m^2 + \frac{\gamma \sqrt{3}\hbar}{2 m^2}}$
and $D \gtrsim \lambda_c \,\, \forall \,\, m$. If $\beta = 4$, $D > \lambda_c$
(it is sufficient that $\beta > 1.26$).
In this way is possible to have a universe dispersed of ultra-light particles ($m\ll m_P$) but
confined inside a sub Planck region and then with a very small cross section.
The limit of this duality is that we can not create such type of ultra-light
black hole because any particle we are able to create in laboratory has $\lambda_c \gg l_P$
where $l_P \propto \sqrt{a_0}$ is the diameter of the throat.
To obtain such ultra-light black hole we should create in the laboratory larger black
hole that subsequently evaporates.
The duality can have a physical relevance also in the case of gravitational
collapse and subsequent evaporation. In this case, because of the evaporation process,
%needs an infinite time,
we can have an evolution toward an ultra-light black hole.
\begin{figure}
 \begin{center}
 \hspace{0.1cm}
  \includegraphics[height=9.5cm]{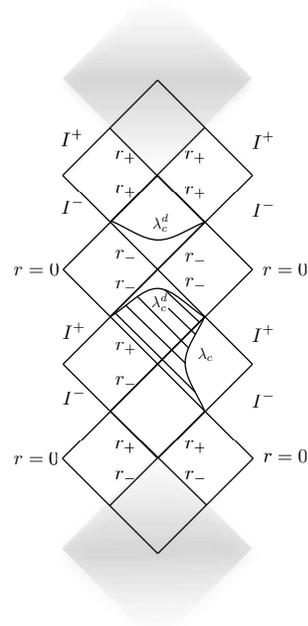}
      \end{center}
  \caption{\label{PBHD2}
Conjectured Carter-Pensrose diagram for the Particle-Black Hole Duality.
The dashed region in the middle of the diagram
represents $\lambda_c^d \leqslant r \leqslant \lambda_c$. It is evident that
a particle in our universe is a black hole for the dual observer.
  }
%\label{temperature}
  \end{figure}

{\em Quantum Duality.}
What we have described in this section is rigorously supported from the results
but we would like to extend
the duality to all the physical particles. In this case we do not have rigourous arguments to support
our speculative idea. This is particularly speculative because we have not examined charged or spinning LQG black holes, and all observed particles have either charge or spin or both. However, if the metric outside a particle with $m < m_P$
and $\lambda_c> 2 G_N m > \sqrt{a_0}$
(in other words an ordinary particle with $\lambda_c$ bigger then its Schwarzschild radius
which is bigger then the radius of the throat)
has the same duality properties of (\ref{metricabella}),
then we can conclude that for any physical particle we have a dual black hole
and the contrary. In fact, if
$\lambda_c^d$ is the Compton wave length seen by the dual observer, defined by
$\lambda_c^d := \hbar/2m_d$, and the dual mass is defined by
$2 G_N m_d = R_+ = a_0/2 m {\mathcal P}^2 G_N$, we obtain that
$\lambda_c^d > r_-$ or $\lambda_c^d < R_+$.
Again, if $\lambda_c> 2 G_N m$ or $m < m_P/2$ the quantum particle is not a black hole
in our universe but it is seen as a black hole
from the perspective of a dual observer if $\lambda_c^d < R_+$ or
$m_d< a_0/4 G_N^2 {\mathcal P}^2 m$. If we parameterise $a_0= \gamma\beta \frac{\sqrt{3}}{2} l_P^2= \gamma\beta \frac{\sqrt{3}}{2} \hbar G_N$
the last condition becomes $m< \gamma\beta \frac{\sqrt{3}}{2} m_P/2 {\mathcal P}^2$ that is always realised for
$m < m_P/2$ if $\gamma\beta \frac{\sqrt{3}}{2}/{\mathcal P}^2 > 1$. In LQG $1\leq\beta\leq 4$ and
${\mathcal P} \approx \gamma^2 \delta_b^2/4$ then
$\gamma \beta \frac{\sqrt{3}}{2}/{\mathcal P}^2 \approx 8 \sqrt{3}/\gamma^3 \delta_b^4 \approx 7.2$ for
$\beta = 1$,
$\gamma \sim 0.2375$
and $\delta_b = 2 \sqrt{3}$.
Under the assumption explained in this section we can conclude
that a particle in our universe is a black hole for a dual observer and vice versa.
When the mass is the range $m_P/2 <  m < \beta \, m_P/2 {\mathcal P}^2$ both the observers
see a black hole.

 \subsection{LQG and LQBHs}
 In this section we want to emphasise the consistency of the metric solution
 with general relativity and full LQG theory.
 The solution reproduces exactly the Schwarzschild solution outside the event horizon and
 it is asymptotically flat, this shows that the metric has the correct semiclassical limit at large
 distance and revels strong deviations from general relativity at the Planck scale.

 The semiclassical solution is also consistent with another result in the full theory \cite{book}.
Historically the idea that the macroscopic geometry emerges
taking the limit of an infinitely dense lattice of loops was  widespread but the result was just the opposite in LQG.
When the density of loop increases the accuracy of the approximation did not increase
but instead the eigenvalue of the area operator increases.
To understand this point we recall the argument.
We consider a weave state that approximates the flat spatial metric
$g^{(0)}_{ab}(x) = \delta_{ab}$ or, in terms of density triad, $e^{(0)i}_a(x) = \delta^i_a$.
We construct a spatially uniform weave state $|s_{\ell_0} \rangle$  formed by
an entanglement of loops of coordinate density $\rho = L/V = 1/\ell_0^2$ ($L =$ total coordinate
length of the loops, $V=$ total coordinate volume). $\ell_0$ represents the average distance
of the loops from each other.
Now we decrease the distance between the loops decreasing the average distance
from $\ell_0$ to $\ell$. We consider the area spectrum of a surface
operator $\hat{A}_S$ in the volume $V$. The averages of the operator $\hat{A}_S$
on the two weave states, $ | s_{(\ell_0)} \rangle$ and $| s_{(\ell)} \rangle$,
are related by the following relation,
\begin{eqnarray}
\langle s_{(\ell)} |\hat{A}_S | s_{(\ell)} \rangle \propto
\frac{\ell_0^2}{\ell^2} \langle s_{(\ell_0)} |\hat{A}_S | s_{(\ell_0)} \rangle.
\label{areaw}
\end{eqnarray}
The result (\ref{areaw}) shows that when the average distance decrees the area increases
and then when we add loops we do not improve the approximation of the metric but
instead we approximate another metric,
\begin{eqnarray}
%\langle s_a |\hat{A}_S | s_a \rangle \sim \frac{a^2}{a_0^2} \langle s_{a_0} |\hat{A}_S | s_{a_0} \rangle.
g_{ab}^{(\ell)}(x) \propto \frac{\ell_0^2}{\ell^2} \, \delta_{ab}. %g_{ab}^{{(\ell_0)}}(x).
\label{metricaa0}
\end{eqnarray}
The physical density of loops, $\rho_{\ell}$, does not change by decreasing $\ell$,
\begin{eqnarray}
\rho_{\ell} = \frac{(\ell_0/ \ell) L}{(\ell_0/\ell)^3 V} = \frac{\ell^2}{\ell_0^2} \rho = \frac{1}{\ell_0^2},
\label{dens}
\end{eqnarray}
and it is natural to identify average distance between the
loops with the Planck length, $\ell_0 \propto l_P$.
\begin{figure}
 \begin{center}
 \hspace{0.1cm}
  \includegraphics[height=4.0cm]{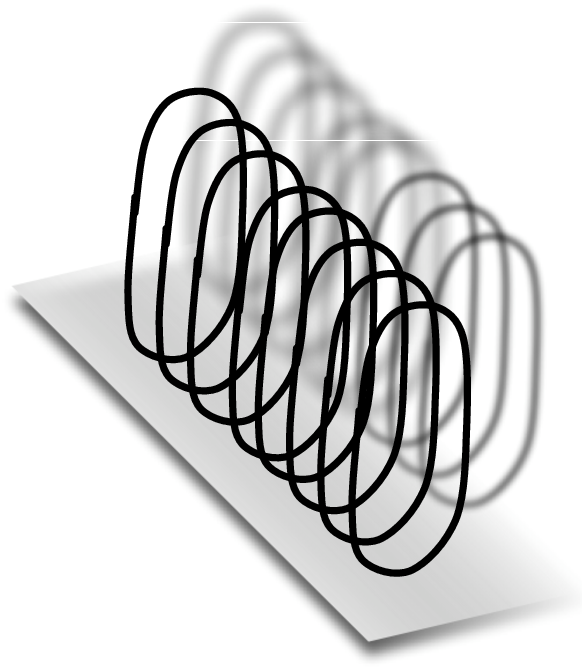}
   \includegraphics[height=4.0cm]{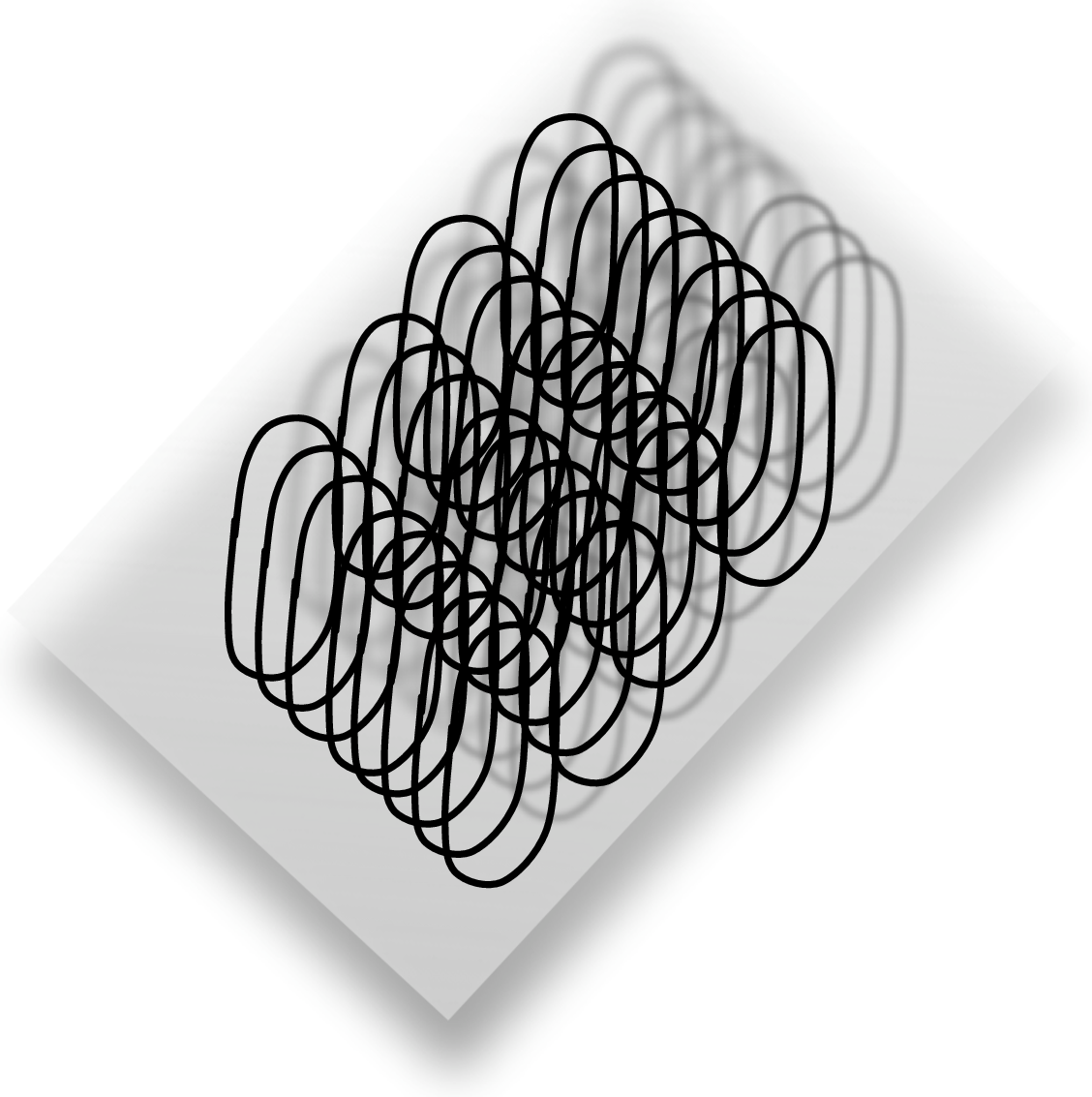}
      \end{center}
  \caption{\label{Weave} The loops weave the space .}
%\label{temperature}
  \end{figure}
The metric in this paper is consistent with the LQG analysis above.
In LQG when we try to probe the substructure beyond the Planck scale
we finish in another bigger geometry. %, analogously i
In LQBH when we
try to go beyond the high curvature Planck scale we finish in another dual
and asymptotically flat classical universe. Comparing the LQBH with the above analysis
in the full theory we can conclude that the bigger geometry discovered in LQG
could describe another classical universe.

\section{Phenomenology}
In this section we study the thermodynamics of LQBH and a possible
interpretation of the dark matter in terms ultra-light LQBH.
We recall the thermodynamics: temperature, entropy and evaporation. % of LQBH and

\subsection{Thermodynamics} \label{LQBHT} %Temperature, entropy and evaporation}
In this section we study the thermodynamics of the LQBH \cite{BR}  \cite{AM}.
The form of the metric calculated in the previous section has the general form,
%\begin{eqnarray}
$ds^2 = - g(r) dt^2 + dr^2/f(r) + h^2(r) (d \theta^2 + \sin^2 \theta d \phi^2)$,
%\label{generalmet}
%\end{eqnarray}
where the functions $f(r)$, $g(r)$ and $h(r)$ depend on the mass parameter $m$
and are the components of the metric (\ref{metricabella}). %given in the table of the first section.
We can introduce the null coordinate $v$ to express the metric
%(\ref{generalmet})
above in the Bardeen form.
The null coordinate $v$ is defined by the relation $v = t + r^{\ast}$, where
$r^{\ast} = \int^r d r/\sqrt{f (r) g(r)}$ and the %exact
differential
is $d v = d t + d r/\sqrt{f(r) g(r)}$. In the new coordinate the metric is,
%\begin{eqnarray}
 $d s^2 = - g(r) d v^2 + 2 \sqrt{g(r)/f(r)} \, d r  dv + h^2(r) d \Omega^{(2)}$.
 %(d \theta^2 + \sin^2 \theta d \phi^2).
 %\label{Bardeen}
%\end{eqnarray}
\noindent

\begin{figure}
 \begin{center}
 \hspace{-0.3cm}
  \includegraphics[height=4.0cm]{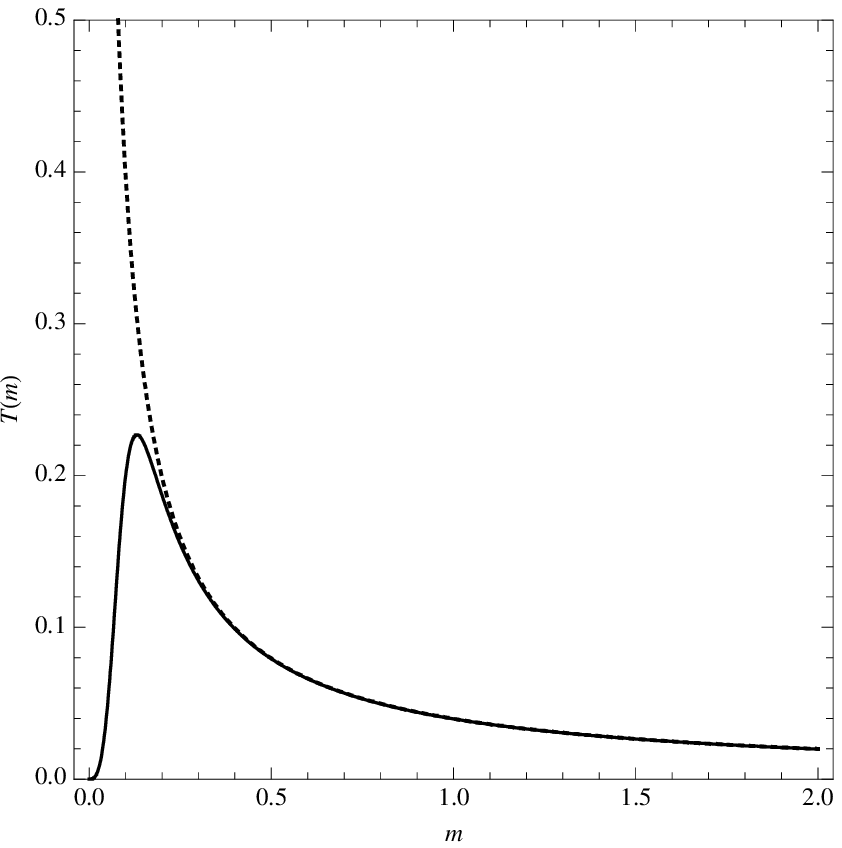}
   \includegraphics[height=4.0cm]{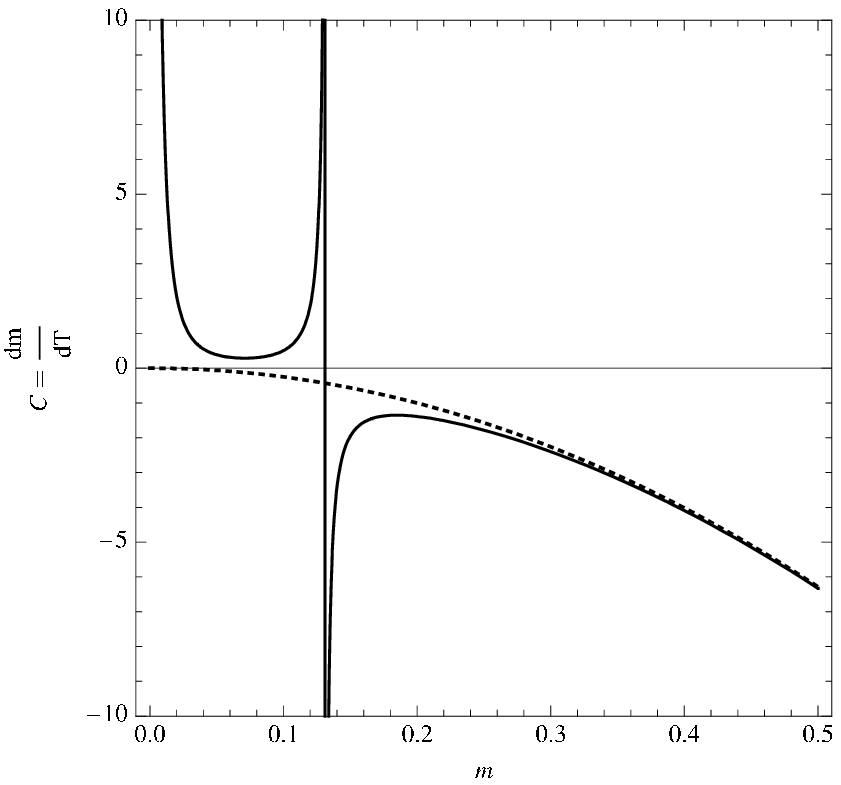}
      \end{center}
  \caption{\label{Pot}
 Plot of the temperature $T(m)$ on the left and of the heat capacity $C_s =\frac{d m}{dT}$ on the right. The continuous plots represent the LQBH quantities, %temperature and
 the dashed lines represent the classical quantities. %Hawking temperature $T=1/8 \pi m$.
  }
\label{temperature}
  \end{figure}

\paragraph{Temperature.}
In this paragraph we calculate the temperature
 for the quantum black hole solution and analyze the evaporation process.
The Bekenstein-Hawking temperature is given in terms of the surface gravity
$\kappa$ by $T= \kappa/2 \pi$, the surface gravity is defined by
%\begin{eqnarray}
$\kappa^2 = - g^{\mu \nu} g_{\rho \sigma} \nabla_{\mu} \chi^{\rho} \nabla_{\nu}
\chi^{\sigma}/2 = - g^{\mu \nu} g_{\rho \sigma}  \Gamma^{\rho}_{\mu 0} \Gamma^{\sigma}_{\nu 0}/2,
$
%\end{eqnarray}
where $\chi^{\mu}=(1,0,0,0)$ is a timelike Killing vector and $\Gamma^{\mu}_{\nu \rho}$
is the connection compatibles with the metric $g_{\mu \nu}$. % of (\ref{generalmet}).
Using the semiclassical metric %in the table of section one
we can calculate the surface gravity
in $r = 2m$ obtaining
%\begin{eqnarray}
%$\kappa^2 = -  g^{00} g^{11} \left(\partial g_{00}/\partial r \right)^2 /4 =
%\left(16 m/64 m^2 + \gamma^2 \delta^2 \right)^2$.
%\end{eqnarray}
%Therefore
and then the temperature, %is
\begin{eqnarray}
T(m) = \frac{128 \pi \sigma(\delta_b) \sqrt{\Omega(\delta_b)} \, m^3}{1024 \pi^2 m^4 + A_{Min}^2},
\label{Temperatura}
\end{eqnarray}
where $\Omega(\delta_b) = 16 (1 + \gamma^2 \delta_b^2)^2/(1 +
     \sqrt{1 + \gamma^2 \delta_b^2})^4$.
The temperature (\ref{Temperatura}) coincides with the Hawking temperature
in the large mass limit.
In Fig.\ref{temperature} we have a plot of the temperature as a function of
the black hole mass $m$.
The dashed trajectory corresponds to the Hawking
temperature and the continuum trajectory corresponds to the semiclassical one.
There is a substantial difference for small values of the mass, in fact
the semiclassical temperature tends to zero and does not diverge for $m\rightarrow 0$.
The temperature is maximum for $m^* = 3^{1/4} \sqrt{A_{Min}}/\sqrt{32 \pi}$
and $T^*= 3^{3/4} \sigma({\delta_b}) \sqrt{\Omega(\delta_b)}/\sqrt{32 \pi A_{Min}}$.
Also this result, as for the curvature invariant, is a quantum gravity effect,
in fact $m^*$ depends only on the Planck area $A_{Min}$.

\begin{figure}
 \begin{center}
  \includegraphics[height=2.6cm]{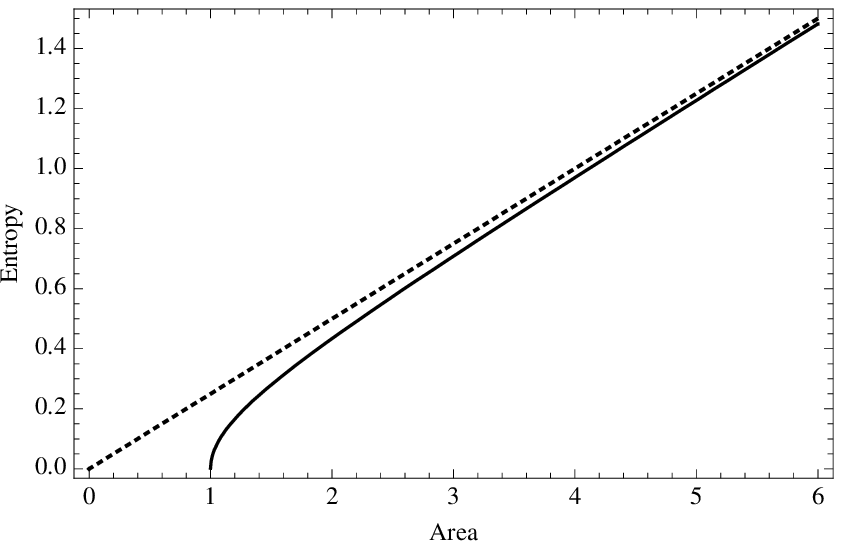}
%  \end{center}
  %\begin{center}
\hspace{0.001cm}
\includegraphics[height=2.62cm]{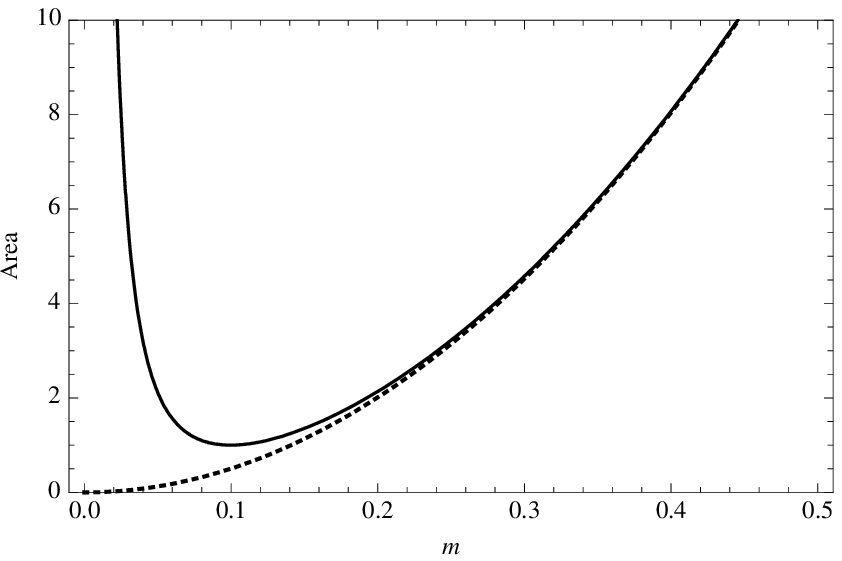}
  %\hspace{1cm}
  \end{center}
  \caption{\label{pbpc2}
  In the first plot we have the entropy for the LQBH as function of the event horizon area
  (dashed line represents
  the classical area low $S_{cl} =A/4$).
  In the second plot we represent the event horizon area as function and the mass
  (dashed line represents
  the classical area $A_{cl}=16 \pi m^2$).}
  \label{Entropy}
  \end{figure}

\paragraph{Entropy.} %Another interesting quantity to calculate is the entropy and its
In this section we calculate the entropy for the LQBH metric.
%quantum corrections.
By definition the entropy as function of the ADM energy is $S_{BH}=\int dm/T(m)$.
Calculating this integral for the LQBH we find
\begin{eqnarray}
S= \frac{1024 \pi^2 m^4 - A_{Min}^2}{256 \pi m^2 \sigma(\delta_b) \sqrt{\Omega(\delta_b)}} + {\rm const.}.
\label{entropym}
\end{eqnarray}
We can express the entropy in terms of the event horizon area.
The event horizon area (in $r=2m$) is
\begin{eqnarray}
A = \int d \phi d \theta \sin \theta \, p_c(r)\Big|_{r = 2m} = 16 \pi m^2 +  \frac{A_{Min}^2}{64 \pi m^2}.
\label{area}
\end{eqnarray}
Inverting (\ref{area})
for $m=m(A)$ and introducing the roots in (\ref{entropym})
we obtain %can express the entropy in terms of the event horizon
%area %and we obtain
\begin{eqnarray}
 S = \pm \frac{\sqrt{A^2 - A_{Min}^2}}{4 \sigma(\delta_b) \sqrt{\Omega(\delta_b)}}   ,
\label{entropyarea}
\end{eqnarray}
where $S$ is positive for $m>\sqrt{a_0}/2$, and negative otherwise. A plot of the entropy is given in Fig.\ref{Entropy}.
The first plot represents entropy as a function of the event horizon area $A$.
The second plot in Fig.\ref{Entropy} represents the event horizon area as function of $m$.
The semiclassical area has a minimum value in $A=A_{Min}$ for
$m= \sqrt{A_{Min}/32 \pi} $. %As for the temperature

We want underline the parameter $\delta_b$ does not play any
regularization rule in the observable quantities $T(m)$, $T^*$, $m^*$ and in the
evaporation process that we will  study in the following section.
We obtain finite quantities taking the limit $\delta_b \rightarrow 0$,
because $\lim_{\delta_b \rightarrow 0} \sigma({\delta_b}) \sqrt{\Omega(\delta_b)}=1$.
%%%%
  %%%
  %%%%
 %%%%
 \paragraph{Evaporation.}
 In this section we focus our attention on the evaporation process
of the black hole mass and in particular in the energy flux from the
hole. First of all the luminosity can be estimated using the
Stefan law and it is given by ${\mathcal L}(m)= \alpha A(m) T_{BH}^4(m)$,
where (for a single massless field with two degree of freedom)
$\alpha = \pi^2/60$, $A(m)$ is the event horizon area and $T(m)$
is the temperature calculated in the previous section.
At the first order in the luminosity the metric above %(\ref{Bardeen})
which incorporates
the decreasing mass is obtained by replacing the mass $m$ with $m(v)$.
% in the metric (\ref{Bardeen}).
%as function of the null coordinate $v$ is also a solution
%but with a new effective stress energy tensor as underlined previously.
Introducing the results (\ref{Temperatura}) and (\ref{area}) of the previous paragraphs
in the luminosity ${\mathcal L}(m)$
we obtain
\begin{eqnarray}
\mathcal{L}(m) =
\frac{4194304 \, m^{10} \pi^3 \alpha \, \sigma^4 \Omega^2}{
(A_{Min}^2+1024\, m^4 \pi^2)^3}.
\label{lumini}
\end{eqnarray}
Using (\ref{lumini}) we can solve the fist order differential equation
\begin{eqnarray}
- \frac{d m(v)}{d v} = \mathcal{L}[m(v)]
\label{flux}
\end{eqnarray}
to obtain the mass function $m(v)$. The result of integration with initial
condition $m(v = 0) = m_0$ is
\begin{eqnarray}
&& \hspace{-0.4cm}  -\frac{n_1 A_{Min}^6+ n_2 A_{Min}^4 m^4 \pi^2+ n_3 A_{Min}^2 m^8
\pi^4- n_4 m^{12} \pi^6}{n_5 m^9 \pi^3 \alpha \,
\sigma(\delta_b)^4 \Omega(\delta_b)^2}  \nonumber \\
&&  \hspace{-0.4cm} +\frac{n_1 A_{Min}^6+ n_2 A_{Min}^4 m^4_0 \pi^2+ n_3 A_{Min}^2 m^8_0
\pi^4- n_4 m^{12}_0 \pi^6}{n_5 m^9_0 \pi^3 \alpha \,
\sigma(\delta_b)^4 \Omega(\delta_b)^2} \nonumber \\
&& \hspace{-0.35cm}= -v,
\label{v(m)}
\end{eqnarray}
where $n_1=5$, $n_2 =27648$, $n_3 =141557760$, $n_4=16106127360$,
$n_5 = 188743680$. From the solution (\ref{v(m)})
we see the mass evaporate in an infinite time. Also in (\ref{v(m)}) we
can take the limit $\delta_b \rightarrow 0$ obtaining a regular quantity
independent from $\delta_b$. In the limit $m \rightarrow 0$ equation (\ref{v(m)})
becomes
\begin{eqnarray}
   \frac{n_1 A_{Min}^6}{n_5  \pi^3 \alpha \,
\sigma(\delta_b)^4 \Omega(\delta_b)^2 \, m^9}   =  v.
\label{v(m)2}
\end{eqnarray}
In the limit $\delta_b \rightarrow 0$, we obtain
$n_1 A_{Min}^6/n_5  \pi^3 \alpha \, m^9 \approx v$. Inverting this equation
for small $m$ we obtain: $m \approx (A_{Min}^6/ \alpha \, v)^{1/9}$.

\subsection{Ultra-light LQBHs as Dark Matter}
It is interesting to consider how the ultra-light LQBHs might manifest themselves if they do exist in nature. Because they are not charged, have no spin, and are extremely light and have a Planck-sized cross-section (due to their Planck-sized wormhole mouth), they will be very weakly interacting and hard to detect. This is especially true as they need not be hot like a light Schwarzschild black hole, but they can be cold as can be seen in Fig.\ref{temperature}. It is thus possible, if they exist, that ultra-light LQBHs are a component of the dark matter. In fact, due to (\ref{Temperatura}), one would expect that all light enough black holes would radiate until their temperature cools to that of the CMB, at which point they would be in thermal equilibrium with the CMB and would be almost impossible to detect. Rewriting (\ref{Temperatura}) for small ${\mathcal P}(\delta_b)$ we get
\begin{eqnarray}
\hspace{-0.5cm} T(m)= \frac{(2m)^3 [1- {\mathcal P}(\delta_b)^2]}{4\pi [(2m)^4+a_0^2]}\approx \frac{(2m)^3}{4\pi [(2m)^4+a_0^2]} .
\label{tempfacile}
\end{eqnarray}

\begin{figure}[tbp]
\begin{equation*}
\leavevmode\hbox{\epsfxsize=8.5 cm
   \epsffile{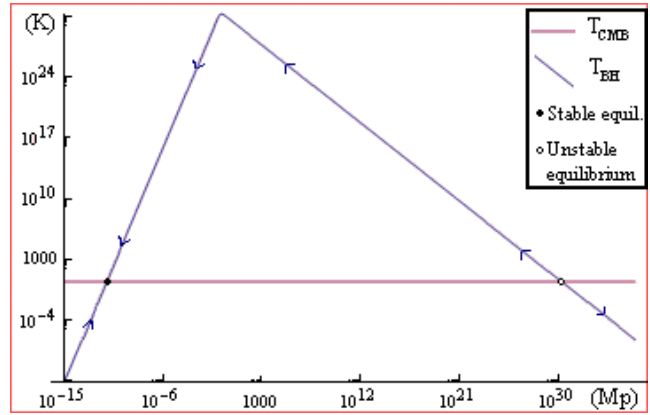}}
\end{equation*}%
\caption{Log-log graph of (\ref{tempfacile}) in units of Kelvin and Planck masses. The constant line denotes the temperature of the CMB radiation; above this line the black hole is hotter than the CMB and so it will lose more energy than it gains, below this line the black hole is colder than the CMB and so it will absorb more CMB radiation than it will emit radiation, thereby gaining mass. The arrows on the temperature curve denote in which direction the black hole will evolve through thermal contact with the CMB.  At the two points where the temperature curve intersects the constant $T_{CMB}$ curve, the black hole is in thermal equilibrium. }
\label{logtempera}
\end{figure}

We thus see emerge a new phenomenon that was not present with Schwarzschild black holes: a black hole in a stable thermal equilibrium with the CMB radiation. In the Schwarzschild scenario, it is of course possible for a black hole to be in equilibrium with the CMB radiation, this happens for a black hole mass of $4.50\times10^{22}$ kg (roughly $60\%$ of the lunar mass). This equilibrium is however not a stable one because for a Schwarzschild black hole the temperature always increases as mass decreases
and vice versa (see the dashed line in Fig.\ref{temperature}, and so if the black hole is a bit lighter than the equilibrium mass it will be a bit hotter than $T_{CMB}$, the temperature of the CMB radiation, and will emit more energy than it gains thus becoming lighter and lighter. Similarly, if the black hole has mass slightly superior to the equilibrium mass, it will be colder than $T_{CMB}$ and thus absorb more energy than it emits, perpetually gaining mass. For the LQBH however, when m is smaller than the critical mass {\sdmass}
of the order of the Planck mass, the relationship is reversed and the temperature increase monotonically with the mass, this allows for a stable thermal equilibrium in this region as is shown in
Fig.\ref{logtempera}.
The values of the black hole mass in the two equilibrium
positions in the LQG case are thus
\bea
&& m_{{\rm unstable}} = 4.50\times10^{22} \, {\rm kg} , \nonumber \\% and}\\
&& m_{{\rm stable}} \approx 10^{-19} \, {\rm kg,} \label{mstable}
\eea
where we have used $\gamma = 0.2375329...$ \cite{gamma} for the Immirzi parameter and assumed $\beta\approx1$. The unstable mass is essentially the same as in the Schwarzschild case while the stable mass, though it formally depends on the value of $\delta_b$, it is quiet insensitive to the exact value of the later as long as $\delta_b$ is at most of the order of unity in which case $m_{{\rm stable}}$ (which is order of magnitude of the mass of the flue virus) is accurate to at least two decimal places.

{\em The following picture thus emerges in LQG}: black holes with a mass smaller than $m_{{\rm stable}}$ grow by absorbing CMB radiation, black holes with a mass larger than $m_{{\rm stable}}$ but smaller than $m_{{\rm unstable}}$ evaporate towards $m_{{\rm stable}}$ and finally black holes with a mass greater than $m_{{\rm unstable}}$ grow by absorbing the CMB radiation.

%Were these ultra-light LQG black hole to make up the majority of the dark matter, (\ref{mstable}) implies that they should have a large-scale number density of approximately $10^{-7}/m^3$. A possible observation to determine whether dark matter is composed in part of ultra-light LQG black hole is to look at regions of the universe which are not in thermal equilibrium, for example after and explosion of some sort, but also contain dark matter. If it can be seen that the dark matter thermalises (as seen from non-conservation of energy of baryonic matter) , but only through radiation, it would be a good indication that ultra-light LQG black holes were part of the dark matter.
%%%%%
%%%%%
%%%%
\subsection{LQBHs Production in the Early Universe}
\begin{figure}
 \begin{center}
 \hspace{-0.3cm}
  \includegraphics[height=4.8cm]{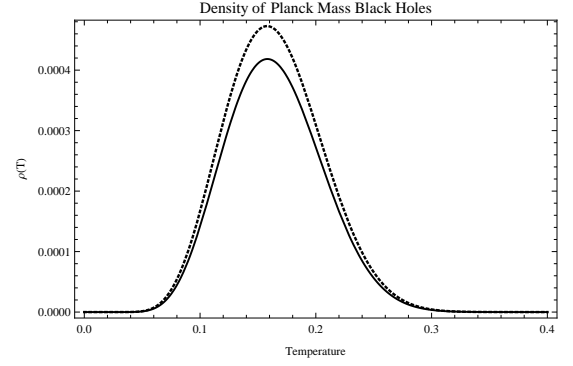}
  \caption{$\rho(T)$, the LQBHs density created due to fluctuations for $m=1$,
%  the Planck mass in Planck units
  and $a_0=0.5$ in Planck units (the value $0.5$ here is chosen to amplify the difference with
  the classical Schwarzschild black hole). The dashed line represents the classical
  density for $a_0=0$.}
  \label{rhoPM}
  \end{center}
  \end{figure}

\begin{figure}
 \begin{center}
  \hspace{-0.15cm}
  \includegraphics[height=5.2cm]{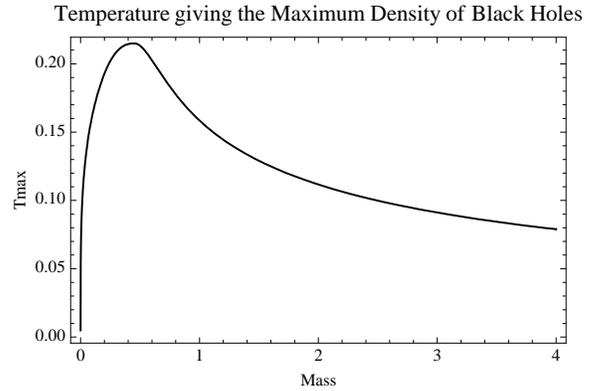}
 \caption{The temperature $T_{Max}$ (Eq.\Ref{Tmax}) at which the density of black holes created through fluctuations is maximised as a function of the mass of the black holes in Planck units. Observe that the temperature is of the order of the Planck temperature $T_P$ in the given mass range. Here we used $\beta=4$.}
  \label{Trhomax}
  \end{center}
  \end{figure}

\begin{figure}
 \begin{center}
  \hspace{-0.3cm}
  \includegraphics[height=5.3cm]{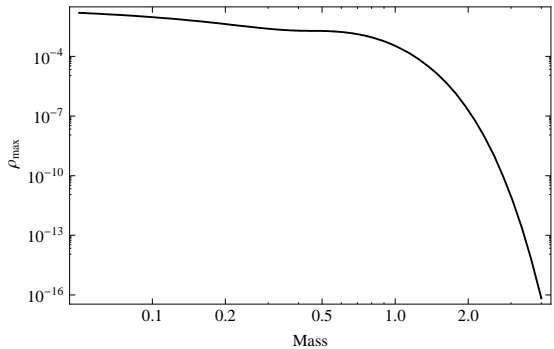}
\caption{The maximal value of
$\rho(m,T) \approx \exp(-\Delta F / T )/\pi^3$
as a function of the mass $m$. The value of the temperature $T$ at which the maximal value of $\rho$ is attained is plotted in Fig.\ref{Trhomax}. Both the mass $m$ and the temperature are in Planck units. Here we used $\beta=4$.}
  \label{maxrho}
  \end{center}
  \end{figure}

\begin{figure}
 \begin{center}
 \hspace{-0.4cm}
  \includegraphics[height=5.1cm]{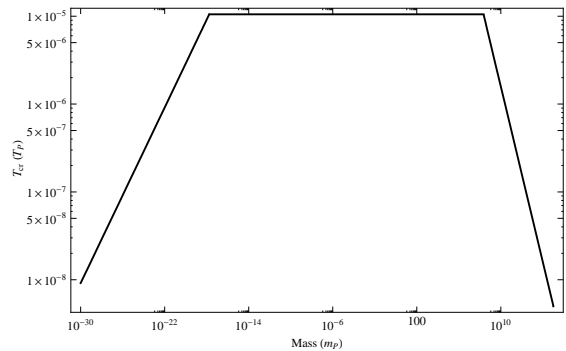}
  %\hspace{1cm}
  \end{center}
  \caption{\label{Trhomaxmod}
  The temperature $T_{cr}$ at which the density of black holes created through fluctuations is maximised as a function of the mass of the black holes in Planck units where we take into consideration the fact that for temperature higher than $T_{eq} \lesssim 10^{15}$GeV black holes do not have time to form because of the rapid expansion. Here we used $\beta=4$ and $T_{eq}= 13\%\times 10^{15}$GeV. We note that for the physically relevant range $10^{-17} \, m_P \leq m \leq 10^{8}\, m_P$, $T_{cr}(m) = T_{eq}$; this is the case for all $T_{eq}$ between $1$\% and 100 \% of $10^{15}$GeV.}
  \end{figure}

\begin{figure}
 \begin{center}
 \hspace{-0.8cm}
  \includegraphics[height=5.6cm]{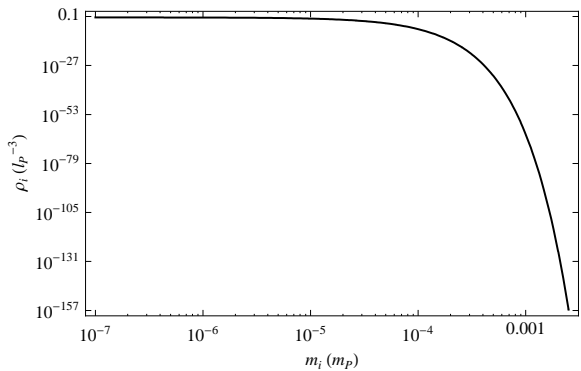}
  %\hspace{1cm}
  \end{center}
 \caption{The initial density of primordial black holes as given by \rf{rhoi} as a function of the initial mass of the black hole. Both the mass $m$ and the temperature are in Planck units. Here we used $\beta=4$ and $T_{eq}= 13\%\times 10^{15}$GeV. The choice of $T_{eq}$ is significant here because the density is very sensitive to $T_{eq}$.}
\label{maxrhomod}
  \end{figure}

%%%%
We can estimate the number of ultra-light LQBHs created as well as the extent of their subsequent evaporation. As exposed in \cite{Kapu}, the probability for for fluctuations to create a black hole is $\exp(-\Delta F / T )$, where $\Delta F$ is the change in the Helmholtz free energy and $T$ is the temperature of the universe. From (\ref{entropym}) and (\ref{tempfacile}) the free energy of a LQBH
 of mass $m$ is
\bea
\hspace{-0.3cm} F_{BH} = m - T_{BH} S_{BH} = m - \half m \, \left[\frac{16 m^4 - a_0^2}{16 m^4 + a_0^2}\right] , \label{Fbh}
\eea
where $T_{BH}$ and $S_{BH}$ are the temperature and entropy of the black hole respectively. The free energy for radiation inside the space where the black hole would form is
\bea
F_{R} = - \frac{\pi^2 \kappa }{45} \, T^4 V,
\label{Fgas}
\eea
where $V$ is the volume inside the 2-sphere which will undergo significant change (i.e. significant departure from the original flatness) in the event of a black hole forming. In the case of a black hole of mass $m\geq \sqrt{a_0}/2$, this is naturally the horizon. Since the horizon has an area of  $4\pi [(2m)^2 + a_0^2/(2m)^2]$, we have that the volume of the flat radiation-filled space in which will undergo the transition to a black hole is $V= (4 \pi/3) [(2m)^2 + a_0^2/(2m)^2]^{3/2}$. However, as we saw earlier, for example in Fig.\Ref{K0} and \Ref{smalllqgbh}, if $m\leq \sqrt{a_0}/2$, a throat of a wormhole forms at $r= \sqrt{a_0}$ and a large departure from flat space is observed. Since the mouth of the worm-hole as area $A_{min}= 8\pi a_0$ we have that the volume of flat space which will be perturbed to create the black hole is $V= (4 \pi/3) (2 a_0)^{3/2}$.
In (\ref{Fgas}) $\kappa$ depends on the number of particles that can be radiated where $\kappa=1$ if only electromagnetic radiation is emitted and $\kappa =36.5$ if all the particles of the Standard Model (including the Higgs) can be radiated.
Hence, if we define
\begin{eqnarray}
\Delta F = F_{BH} - F_{R}
\end{eqnarray}
to be the difference between the black hole free energy and the radiation free energy inside the volume which is to be transformed, we have, in Planck units, that the density of black holes of mass $m$ coming from fluctuations is of the order of
\bea
\rho(m) \approx \frac{1}{\pi^3}\exp(-\Delta F / T ). \label{deltaq}
\eea
Plotting $\rho$ as a function of $T$, (see for example Fig.\ref{rhoPM}) we see that $\rho$ peaks at a given temperature which %as can be seen from Fig.(\ref{Trhomax})
is of order $T_P$ for a black hole mass of order $m_P$. If we imagine that the universe started in a hot Big Bang and gradually cooled, looking at Fig.\Ref{rhoPM}, we see that at very high temperatures the amount of black holes of a given mass created from fluctuations is relatively small. Then as the universe starts to cool, the number of black holes increases until it reaches a maximum value at $T_{Max}(m)$ (see Fig.\ref{Trhomax}) at which point, when the universe cools further, no more black holes of mass $m$ are created and the existing black holes start to evaporate.
By varying \Ref{deltaq} with respect to $T$, we find that $T_{Max}(m)$ the temperature for which the maximum amount of black holes are formed is
\begin{equation}
%\hspace{0.0cm}
T_{Max}(m) \hspace{-0.07cm} = \hspace{-0.07cm}
\begin{cases} \frac{\sqrt{3} 5^{1/4} \left(m \left(3 a_0^2+16 m^4\right)\right)^{1/4}}{ 2^{9/8} \left({a_0}^{3/2} \left(a_0^2+16 m^4\right) \kappa \right)^{1/4} \pi^{3/4}} &
\hspace{-0.2cm}
\text{if $m\leq \sqrt{a_0}/2$,}
\\
\frac{\sqrt{3} m 5^{1/4}\left(3 a_0^2+16 m^4\right)^{1/4}}{\left(a_0^2+16 m^4\right)^{5/8} \kappa^{1/4}\pi^{3/4}} &
\hspace{-0.2cm}
\text{if $m\geq \sqrt{a_0}/2$.} \end{cases}
\label{Tmax}
\end{equation}
Combining \Ref{deltaq} and \Ref{Tmax}, we can obtain the maximal primordial density of black holes
$\rho_{max}$. Fig.\ref{maxrho} is a graph of this quantity in Planck units and for $\beta=4$.
%
%
%
%\begin{figure}[tbp]
%\begin{equation*}
%\leavevmode\hbox{\epsfxsize=8.5 cm
 %  \epsffile{rhopmt.eps}}
% \epsfile{rhoT.eps}}
%\end{equation*}%
%\caption{$\rho(T)$, the LQBHs density created due to fluctuations for $m=m_P=1$, the Planck mass in Planck units and %a_0=0.5$ in Planck units. }
%\label{rhoPM}
%\end{figure}
%
%
%\begin{figure}[tbp]
%\begin{equation*}
%\leavevmode\hbox{\epsfxsize=8.0 cm
%   \epsffile{Trhomax.eps}}
%\end{equation*}%
%\caption{The temperature $T_{Max}$ (Eq.\Ref{Tmax}) at which the density of black holes created through fluctuations is maximised as a function of the mass of the black holes in Planck units. Observe that the temperature is of the order of the Planck temperature $T_P$ in the given mass range. Here we used $\beta=4$.}
%label{Trhomax}
%\end{figure}
%
One more subtlety however must be considered the number of black holes produced can be calculated. Formula \rf{deltaq} is only valid if the universe can reach local equilibrium. If the time scale for the expansion of the universe is much shorter than the time scale for collisions between the particles, the universe expands before equilibrium can take place and so \rf{deltaq}, which requires equilibrium, is not valid. It can be shown \cite{mukhanov}, that local equilibrium is reached for temperatures
\begin{align}
T\ll 10^{15} {\rm GeV}-10^{17}{\rm GeV}.\label{teqcond}
 \end{align}
 This means that before the universe cooled down to temperatures below $10^{15}$GeV, the universe expanded too quickly to have time to create black holes from fluctuations in the matter density. The fact that the universe must first cool down to below $10^{15}$GeV before a black holes can be created means that black holes of mass $m$ will not be created at temperature $T_{Max}(m)$ of \rf{Tmax} but rather at temperature $T_{cr}(m)=\min\{T_{Max}(m),T_{eq}\}$ where $T_{eq} \lesssim %\underset{\sim}{<}
 10^{15}$GeV is the temperature below which local equilibrium can be achieved and thus black holes can be created. As can be seen in Fig.\ref{Trhomaxmod}, this means that for a significant range of black hole masses, from about $10^{-17} \, m_P$ to $10^{9} \, m_P$ the maximal density will be created when the universe reaches temperature $T_{eq}$. As it turns out, this range will encompass the quasi-totality of black holes responsible for dark matter or any other physical phenomenon considered in this paper.
 %
%\begin{figure}[tbp]
%\begin{equation*}
%\leavevmode\hbox{\epsfxsize=8.5 cm
%  \epsffile{Trhomaxmod.eps}}
% \epsffile{Prova.eps}}
%\end{equation*}%
%\caption{The temperature $T_{cr}$ at which the density of black holes created through fluctuations is maximised as a function of the mass of the black holes in Planck units where we take into consideration the fact that for temperature higher than $T_{eq} \lesssim 10^{15}$GeV black holes do not have time to form because of the rapid expansion. Here we used $\beta=4$ and $T_{eq}= 4.3\%\times 10^{15}$GeV. We note that for the physically relevant range $10^{-20} \, m_P\leq m\leq 10^{10}\, m_P$, $T_{cr}(m) = T_{eq}$; this is the case for all $T_{eq}$ between $1$\% and 100 \% of $10^{15}$GeV.} %5 $5$
%\label{Trhomaxmod}
%\end{figure}
The fact that black holes are created only once the universe has cooled down to $T_{eq}$ entails that the initial density of black holes is
\begin{eqnarray}
\rho_{i}(m) \approx \frac{1}{\pi^3}\exp(-\Delta F(m) / T_{cr}(m) ), \label{rhoi}
\end{eqnarray}
(where the dependencies on the black hole mass $m$ are explicitly written) and not of the density plotted in Fig.\ref{maxrho}. Graphing \rf{rhoi}, we see in Fig.\ref{maxrhomod} that only black holes with an initial mass of less then $10^{-3} \, m_P$ are created in any significant numbers.

{\em We are thus presented with the following picture}: as the temperature cools from the Big Bang, and the expansion of the universe starts to slow down fluctuations of the matter start producing Ultra-light black holes of a thousandths of the Planck mass and less, as can be seen from Fig.\ref{maxrhomod}. Once the this initial density of black holes is formed and the universe start cooling further, the primordial black holes will start to evaporate since they will be hotter than the surrounding matter.

\subsection{Evaporation of Ultra-light LQBHs}

\begin{figure}[tbp]
\begin{equation*}
\leavevmode\hbox{\epsfxsize=8.5 cm
   \epsffile{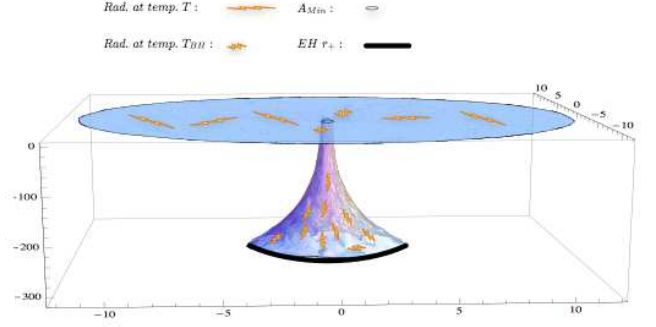}}
\end{equation*}%
\caption{The black hole horizon and its accompanying patch of space are in thermal equilibrium at temperature $T_{BH}$. The rest of the universe has radiation in thermal equilibrium at temperature $T$. The two can interact radiatively through a Planckian surface of area $A_{min}$.}
\label{entonnoir}
\end{figure}

\begin{figure}[tbp]
\begin{equation*}
\hspace{-0.5cm}
\leavevmode\hbox{\epsfxsize=8.0 cm
   \epsffile{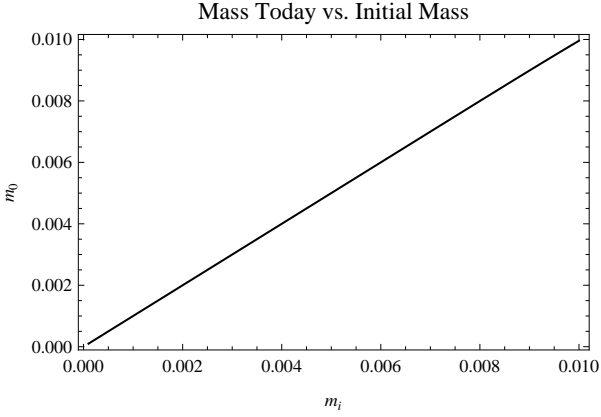}}
\end{equation*}%
\caption{The mass today $m_0$ of a black hole created with mass $m_i$ during the Big Bang. Both masses are in Planck units. $m_0$ is obtained from $m_i$ through Eq.\Ref{evapanorm} ($\beta = 4$).}
\label{mzero}
\end{figure}

Once the black holes are formed, the only way they can disappear is through evaporation. When the mass, $m$, of a black hole satisfies $m \geq \sdmass$, the LQBHs evaporate like a Schwarzschild black holes would:  %planck
\bea
\frac{d m}{d t} = \frac{\pi^2}{60}A(m)T^4 - \frac{\pi^2}{60}A(m)(T_{BH}(m))^4, \label{evapnorm}
\eea
where $\pi^2/60$ is Stefan-Boltzmann's constant in Planck units, $A(m)$ is the area of the LQBH horizon, $T$ is the temperature of the radiation in the universe and $T_{BH}(m)$ is the temperature of the LQBH. So the first term in the last equation represents the radiation absorbed by the black hole while the second term is the radiation emitted by the black hole. Things take on a new twist however when the mass falls below \sdmass, which will happen within 1000 years of the Big Bang for black holes created with an initial mass of less than $100 \, m_P$. As illustrated in Fig.\ref{smalllqgbh}a the black hole horizon as well as the space surrounding it, is separated from the rest of the universe by a wormhole of Planckian diameter. The wormhole as well the chunk of space surrounding the horizon form very slowly and gradually as can be seen from (\ref{v(m)},\ref{v(m)2}). So we can divide space in three parts: 1) the inside of the black hole, 2) a relatively small (compared to the rest of the universe) bag of space in between the black hole horizon and the mouth of the worm-hole and 3) (infinite) flat space outside of the mouth of the worm-hole. Theoretically, these three subsystems could be at three different temperatures. However, because the size of the horizon of the black hole is greater than the size of the mouth of the worm-hole ($4\pi (2m)^2 + a_0^2/(2m)^2 > 4\pi (2 a_0)$) and becomes ever more so as the black hole gets smaller, the bag of space between the horizon and the mouth of the worm-hole, will thermalise faster with the black hole than with the flat space. Since also the bag starts out with a very small volume and this volume changes only slowly, the thermalisation with the black hole happens rather rapidly (on cosmological scales). Hence, for cosmological purposes, we can suppose that the black hole and the bag of radiation between the horizon and the mouth of the worm-hole are in thermal equilibrium at the temperature of the black-hole, $T_{BH}$, and that the combined system interaction by thermal radiation with the outside flat space through the Planck-sized mouth of the worm-hole which has area $A_{min}$.  We shall label the temperature of the radiation in the flat space (the CMB) $T$. This situation is illustrated in Fig.\ref{entonnoir}.
The volume of the bag of space between the horizon and the mouth of the wormhole is
\bea
&& V_{bag}  =  \int_{r= 2m}^{\sqrt{a_0}} 4\pi g_{\theta \theta} \sqrt{g_{r r}} dr \nonumber \\
   && \hspace{-1cm} \approx   \frac{8 a_0^{3/2} \sqrt{a_0-4 \sqrt{a_0} \, m} \left(a_0+2 \sqrt{a_0} \, m+6 m^2\right) \pi }{15 m^3},\label{vbag}
\eea
if $\delta_b$ is of the order of unity or less (which is the natural choice), where $1<\chi(m)<2^{\frac{3}{2}}$. As it turns out though, the worm-hole radiation term is not at all significant for the value and precision considered here, however, for completeness we will include it.  The energy density of thermal radiation at temperature $T$ is $\pi^2 T^4/15$. Thus the energy of the combined black hole and bag of space in thermal equilibrium with it between the horizon and the mouth of the worm-hole is
$m + \pi^2 \, V_{bag} \, T_{BH}^4/15$. Writing the conservation of energy considering that the two systems (LQBH+ bag and flat space) interact via black body radiation through the mouth of the worm-hole, we get:
\begin{eqnarray}
&& \hspace{-0.5cm}
\frac{d}{d t}\left[m + \left(V_{bag}(m) \, \frac{\pi^2}{15} (T_{BH}(m))^4 \right)\right] \nonumber \\
&& = \frac{\pi^2}{60}A_{min}T(t)^4 - \frac{\pi^2 \kappa}{60}A_{min}(T_{BH}(m))^4 , \label{evapanorm}
\end{eqnarray}
where possible curvature corrections have been neglected. Where we have used that the power of the thermal radiation (in Planck units) emitted by a black body is of surface area $A$ and temperature $T$ obeys the Stefan-Bolzmann law:
 \begin{align}
P = \frac{\pi^2 \kappa}{60} A \, T^4 , \label{stefan}
 \end{align}
 where $\kappa$ depends on the number of particles that can be radiated where $\kappa=1$ if only electromagnetic radiation is emitted and $\kappa =36.5$ if all the particles of the Standard Model (including the Higgs) can be radiated. As we will be dealing with extremely hot temperatures at which all the Standard Model particles are relativistic and thus all particles can be emitted, we will be using $\kappa=36.5$ in what follows though in fact it will make no difference whether we use $\kappa=1$ or $\kappa=36.5$.
 Using (\ref{tempfacile}, \ref{vbag}) and approximating $T(t)\approx T_{CMB} (t_0/t )^{2/3}$, where $T_{CMB}$ is the temperature of the cosmic microwave background today and $t_0$ is the age of the universe. We can make this simplification because this is the equation for the temperature of radiation in a matter dominated universe, and the length of time for which the universe was not matter dominated is negligible in standard cosmology for our purposes. This allows us to calculate the masses of the ultra-light black holes today numerically. We find that, all black holes which initially started
with mass $m_i = 0.001 \, m_P$ are de facto stable: the difference
between the initial mass $m_i$ and the mass of the black hole today $m_0$ satisfies in fact if
\[\frac{m_i - m_0}{m_i} \approx 10^{-14}\]
where we have taken $\beta =4$ (but the result is not sensitive to the exact value of $\beta$) and for smaller initial masses the difference is even smaller.
In Fig.\ref{mzero} are represented different value of the mass $m_0$ of a LQBH today as a function of it's initial mass $m_i$. %In this plot is not introduced any limit to the temperature of the early universe.
%in the range $0.03 \leq m_i \leq 100 m_P$ will have evaporated to the current mass of $m_0 = (0.0183 \pm 0.001) m_P$ if we suppose the polymeric parameter $\delta_b$ to be of order 1 and $\beta=4$. This is illustrated in Fig. \ref{mzero} which shows the mass $m_0$ of a LQBH today as a function of it's initial mass $m_i$.

If, for example, we consider a
LQBH of mass $m_0 = 0.000635 \, m_P$, by Wien's Law they radiate with maximum intensity at
\begin{eqnarray}
\hspace{-0.3cm} E_{\gamma} = 2 \pi \, m_P \left( \frac{\omega_b}{l_P T_P} \frac{1}{T_{BH}(m)}\right)^{-1} \approx 1.46 \times 10^{19} {\rm eV}.
\label{energyphoto}
\end{eqnarray}
Where $\omega_b = 2.897768551 \times 10^{-3} \, {\rm m K}$, $m_P$ is the Planck mass in eV's and $T_P$ is the Planck temperature in Kelvins.
This %is obviously very hot and
means that the ultra-light black holes would not have had time, in the life-time of the universe, to thermalise with the CMB.
This does not stop them from being very stable in any case
as the calculated value of $m_0$ above shows.
The mass $m_0$ in ${\rm eV}$ is $m_0 \approx  7.75 \times 10^{24} \, {\rm eV}$ and the temperature
in Kelvin degree is $T(m_0) \approx \, 3.44 \times 10^{22} \, {\rm K}$.
%, in fact, if we let these black holes evaporate for another 13-14 billion years they will only lose $8\%$ of their mass and have the almost unchanged mass of $0.0168 m_P$ in the end.
%In Fig.\ref{mzero} are represented different value of the mass $m_0$ of a LQBH today as a function of it's initial mass $m_i$. %In this plot is not introduced any limit to the temperature of the early universe.
%%%
\subsection{%\hspace{-0.24cm}
Number of e-folds Elapsed Since LQBHs \\ \hspace{0.2cm}Creation  to Account for Dark Matter}

For all black hole initial mass $m_i$, we know, thanks to Eq.\Ref{evapanorm} what the black hole's current mass is. We also know what the initial concentration of each type of black hole was
from Eq.\Ref{rhoi}. In addition, we know that the current matter density for dark matter is approximately $0.22\rho_{crit}$. If we now suppose that currently, all dark matter is actually composed of ultra-light black holes, we have that
\begin{align}
\int_{0}^\infty\frac{(a(t_i))^3 m_0(m_i) \rho_{max}(m_i)}{(a(t_0))^3}d m_i = 0.22\rho_{crit},\label{expander}
\end{align}
where $a(t_0)$ is the scale factor of the Universe at present ($t_0$), $a(t_i)$ is the scale factor of the universe when the primordial black holes were created ($t_i$) and so $\frac{(a(t_i))^3 \rho_{i}(m_i)}{(a(t_0))^3}$ is the current number density of black holes of mass $m_0(m_i)$. Since the scale factor does not depend on $m_i$, we can rearrange this equation to find out the number of e-folds $N_e$ that the universe is required to have expanded since the creation of the primordial black holes for the light black holes to form the totality of dark matter:
\begin{align}
N_e := \ln \frac{a(t_o)}{a(t_i)} = \frac{1}{3}\ln\left(\frac{\int_{0}^\infty\ m_0(m_i) \rho_{i}(m_i) d m_i}{0.22\rho_{crit}}\right), \label{efolds}
\end{align}
and
\begin{eqnarray}
\frac{a_o}{a_i} := \frac{a(t_o)}{a(t_i)} = \left(\frac{\int_{0}^\infty\ m_0(m_i) \rho_{i}(m_i) d m_i}{0.22\rho_{crit}}\right)^{\frac{1}{3}} . \label{aaa}
\end{eqnarray}
The integral in Eq.\Ref{efolds}, is evaluated to give $1.58\times 10^{-12} \, m_P \, l_P^{-3}$.
%$8.5\times 10^{-8} \, m_P \, l_P^{-3}$.
This implies a number of e-folds between the creation of the black holes and the present day of
\begin{eqnarray}
 N_e  \approx  85 \,\,\, \text{ and } \,\,\, \label{NE} %\nonumber \\
 \frac{a_0}{a_i} \approx 10^{37},\label{atheo}
\end{eqnarray}
where we have used $T_{eq}= 1.3\times 10^{14}$GeV and $\beta=4$ though these last two results are very robust under changes of $T_{eq}$ and $\beta$.
\begin{figure}%[tbp]
%\begin{equation*}
\begin{center}
\hspace{-0.7cm}
 \includegraphics[height=5.0cm]{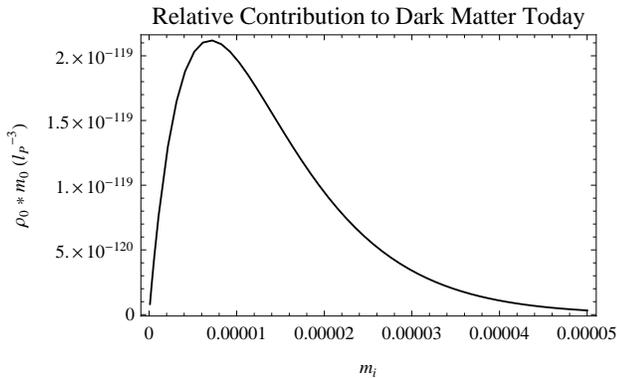}
 \end{center}
%\leavevmode\hbox{\epsfxsize=8.5 cm
%   \epsffile{rhome.eps}}
%\end{equation*}%
\caption{This graph shows the current mass density of black holes as a function of their initial mass $m_i$. $\rho_0 (m_0)$ is the current number density of black holes of mass $m_0$, so
$\rho_0=\rho_{i} \, (a_i^3/a_0^3)$. Because, for all practical purposes, $m_0=m_i$, the area under the curve is the present matter density due to LQBH. If that density is equal to $0.22\rho_{crit}$, the LQBH will account for all dark matter.  From this graph, we see that at present times, LQBH mass density is entirely dominated by black holes which had an initial mass of about $ 10^{-5} m_P$. In this graph we have used $\beta=4$ (the graph is not very sensitive to this choice) and $T_{eq}= 13\%\times 10^{15}$GeV (the numerical values of the graph vertical axis are sensitive to this value but location of the peak and the general shape of the graph are not).}
\label{rhome}
\end{figure}

Thus, if we want all dark matter to be explained by ultra-light black holes, then the universe must have expanded by a scale factor of $10^{37}$ between the creation of the black holes and the present day to achieve an ultra-light black-hole mass density of approximately $0.22 \rho_{crit}$, the estimated dark matter density. Since the end of inflation, the universe has expanded by a scale factor of about $10^{28}$. This implies that the ultra-light black holes have to be created towards the end of the period of inflation which means that inflation should be going on when the universe is at temperature of the order of $10^{14}$GeV$-10^{15}$GeV, this is indeed close to the range of temperatures at which inflation is predicted to happen in the simplest models of inflation ($10^{15}$GeV$-10^{16}$GeV).

 %It is however consistent with Loop Quantum Cosmology which predicts super inflation when the universe is denser than 0.41 Planck density and normal inflation thereafter\cite{yiling}. However, since we already concluded that ultra-light LQBHs could not form the majority of dark matter, the universe should then have expanded much more than 92 e-folds between the production of the black holes and the present day. This leaves open the possibility that the black holes were produced before inflation, allowing inflation to happen some $10^{-36}-10^{-34} s$ after the Big Bang as is usually assumed.

{\em So if black hole make up the majority of dark matter we have the following picture}. Primordial black holes were created during an inflationary period when the universe had a temperature in the $10^{14}$GeV$-10^{15}$GeV range. Since their creation the Universe has expanded by 85 e-folds. From
Fig.\ref{rhome} we see that the majority of the black holes making up the dark matter would have been created with an initial mass of around $10^{-5}$$m_P$; Eq.\Ref{evapanorm} then implies that their mass has changed by less than 1 part in $10^{-14}$ since their creation making these black holes very stable.  That is the case (due to the Planck-sized area of the mouth of the worm-hole) even though the radiation they emit is still very hot. From Wien's law we have that the maximum intensity of their radiation is for particles of energy of about $10^{13}$eV.

\subsection{LQBHs as Sources for \\ Ultra-Hight Energy Cosmic Rays}
Hot ultra-light black holes are very interesting phenomenologically because there is a chance we could detect their presence if they are in sufficient quantities. %Photons of $10^23 eV$ that the ultra-light black holes are extremely energetic and very few sources can emit such highly energetic photons, thus seeing such photons would be good experimental support for ultra-light LQBH as dark matter.
The mass of ultra-light LQBHs today is $m_0 \approx 10^{24} {\rm eV}$, then we can have emission
of  cosmic rays from those object in our galaxy.

In fact, Greisen Zatsepin and Kuzmin proved that cosmic rays which have travelled over $50\, {\rm Mpc}$ will have an energy less than $6\times 10^{19} {\rm eV}$ (called the GZK cutoff) because they will have dissipated their energy by interacting with the cosmic microwave background \cite{GZK}. However, collaborations like HiRes or Auger \cite{HiRes} have observed cosmic rays with energies higher than the GZK cutoff, ultra high energy cosmic rays (UHECR).  The logical conclusion is then that within a
$50 \, {\rm Mpc}$ radius from us, there is a source of UHECR. The problem is that we do not see any possible sources for these UHECR within a $50 \, {\rm Mpc}$ radius. The ultra-light LQBH which we suggest could be dark matter do however emit UHECR. Could it be that these black holes not only constitute dark matter but are also the source for UHECR? This is not such a new idea, it has already been proposed that dark matter could be the source for the observed UHECR \cite{Xpart}.

Let us compare the predicted emissions of UHECR from LQBHs with the observed quantity of UHECR. Detectors of UHECR, like Auger or HiRes, cover a certain surface area $A_D$ and register events of UHECRs hitting their detector. Let us suppose that the source for UHECR is indeed the dark matter. It is believed that dark matter forms an halo (a ball) around the Milky Way of roughly the size of the Milky Way, let $R_{MW}$ be the radius of the Milky Way. We suppose the dark matter is centred in the halo of the Milky Way. $R_{MW}$ is then roughly $50000 \, {\rm ly}$ (light-years). For the purpose of the following calculations, we can suppose that the Earth is on the outer edge of the Milky Way
(in fact it is $30 000 \, {\rm ly}$ from the centre). If we then suppose that all the UHECRs we observe come from the matter halo of the Milky Way, and if the production rate (in particle of UHECR per metre cubed per second) of UHECR is $\sigma$ ($[\sigma] = {\rm particles} \,\, {\rm s}^{-1} {\rm m}^{-3}$),
then we have the halo produces $\frac{4\pi}{3}R_{MW}^3 \sigma$ particles of UHECR per second. Since the Milky Way in equilibrium, that means that $\frac{4\pi}{3}R_{MW}^3 \sigma$ particles of UHECR per second cross the $2$-sphere of area $4\pi R_{MW}^2$ enveloping the Milky Way and its halo.
Thus, with a detector of area $A_D$ on this $2$-sphere, the detector should have a rate of detection of UHECR events of
\begin{align}
R_E = \frac{A_D}{4 \pi R_{MW}^2 }\frac{4\pi}{3}R_{MW}^3 \sigma =  \frac{A_D R_{MW}\sigma}{3} .
\end{align}
Therefore we should have that
\begin{align}
\sigma =  \frac{3 R_E}{A_D R_{MW}} . \label{sigmundfreud}
\end{align}
Let us use Auger's data \cite{HiRes}, for Auger we have that $A_D = 3000 \, {\rm km}^2$ and $R_E = 3$ events per year. This gives us an observed $\sigma$ of
\begin{align}
\sigma_{obs} \approx 10^{-37} \, \frac{ \text{UHECR particles}}{ {\rm s} \,\,  {\rm m}^{3}}.\label{observ}
\end{align}

\begin{figure}
 \begin{center}
 \hspace{-0.15cm}
  \includegraphics[height=5.5cm]{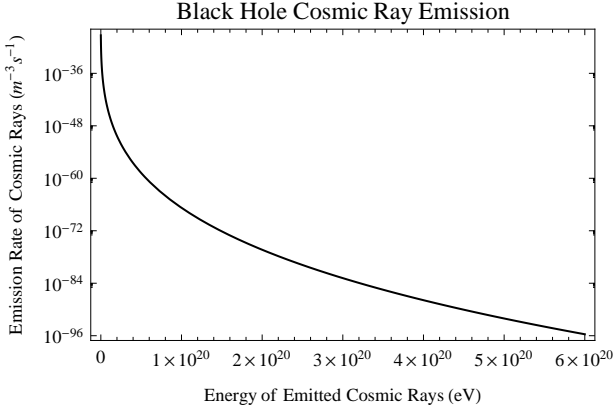}
  %\hspace{1cm}
  \end{center}
  \caption{The average emission rate of particles by primordial ultra light black holes in the universe given by \rf{averageemit} assuming $\beta=4$ and $T_{eq}=1.3\times10^{14}$GeV.}
\label{averageuhecr}
  \end{figure}

\begin{figure}
 \begin{center}
 \hspace{-0.15cm}
  \includegraphics[height=5.6cm]{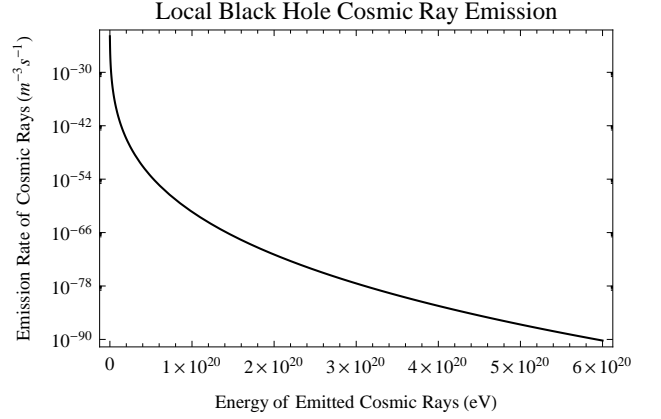}
  %\hspace{1cm}
  \end{center}
  \caption{The local emission rate of particles by primordial ultra light black holes in the Milky Way given by \rf{localemit} assuming $\beta=4$ and $T_{eq}=1.3\times10^{14}$GeV.}
\label{localuhecr}
  \end{figure}

We must compare this value with the predicted production of UHECR by LQBHs.
Using Planck's Law, Eq.\Ref{tempfacile} and the fact that the bag is in equilibrium with the black hole and the pair radiates through the worm-hole mouth of area $A_{min}= 8\pi a_0$ we have that (in Planck units), the rate of emission of particles of energy
$\nu$ by an ultra-light black hole is
\begin{align}
R_{BH}(\nu,m_0) = \frac{2 A_{min}}{\pi} \frac{ \nu^2}{e^{\frac{  \nu }{T_{BH}(m_0)}}-1} .
\end{align}
This implies that the collective rate of emission of particles of energy $\nu$ by all primordial ultra light black holes, on average in the universe, is
\begin{align}
R_{BH}(\nu) = \int_{m_0=0}^{\sqrt{a_0}/2}\rho_0(m_0)\frac{2 A_{min}}{\pi} \frac{ \nu^2}{e^{\frac{  \nu }{T_{BH}(m_0)}}-1} d m_0 ,\label{averageemit}
\end{align}
where $\rho_0(m_0)=\rho_{i}(m_0) (a_i^3/a_0^3)$ is the present day number density of black holes of mass $m_0$. \rf{averageemit} is plotted in Fig.\ref{averageuhecr}.
However, the local dark matter density is much larger than the average dark matter density in the universe. Hence there should be more radiation emitted in our local neighbourhood than on average in the universe.
That the dark matter density of the Milky Way halo, determined by the rotation curves, is calculated to be $\rho_{MWDM}= 0.3 \, {\rm GeV} {\rm cm}^{-3}$\cite{jacob}.
%We have already determined that the LQBHs that would make up dark matter would be almost exclusively black holes of mass $m_0 = 0.000635 \, m_P$, thus the number density of LQBHs i\includegraphics[]{Trhomaxmod.eps}
%s $\rho_{MWDM}/m_0$.
If we suppose that the distribution of ultra-light black holes in the Milky Way is the same as in the universe as a whole, we then have that
\begin{align}
\rho_{MWBH}(m_0) =
\frac{\rho_{MWDM} \, \rho_i(m_0)}{\int_{m=0}^{\infty} \rho_i(m) \, m \, dm} ,\label{milkyrho}
\end{align}
where $\rho_{MWDM}(m_0)$ is the number density of black holes of mass $m_0$ in the Milky Way at present. In this case, analogously to \rf{averageemit}, we have that locally, the collective rate of emission of particles of energy $\nu$ by all local primordial ultra light black holes is
\begin{eqnarray}
&& R_{Local BH}(\nu) = \int_{m_0=0}^{\sqrt{a_0}/2}\rho_{MWBH}(m_0)\frac{2 A_{min}}{\pi} \nonumber \\
&& \hspace{3cm} \times \frac{ \nu^2}{e^{\frac{  \nu }{T_{BH}(m_0)}}-1} d m_0 ,\label{localemit}
\end{eqnarray}
which is plotted in Fig.\ref{localuhecr}.
%%%
%%%
This implies a theoretical production rate of UHECR photons in the Milky Way of
\begin{eqnarray}
 \sigma_{th} = \int_{m_0=0}^{\infty}\int_{6\times 10^{19} \, {\rm eV}}^{m_0} \hspace{-0.2cm}
 \frac{2 A_{min} \, \rho_{MWBH}(m_0) \, \nu^2}{\pi (e^{\frac{  \nu }{T_{BH}(m_0)}}-1)} d \nu .\label{milkproducts}
\end{eqnarray}

As it turns out, the result of $\sigma_{th}$ is very robust for parameters except for $T_{eq}$ on which $\sigma_{th}$ is very sensitive. In order to agree with \rf{observ}, we must have $T_{eq}\approx 13\% \times 10^{15}$GeV. This is in great accordance with \rf{teqcond}. If $T_{eq}\gg 13\% \times 10^{15}$GeV, then ultra light black holes cannot form the majority of dark matter, because if they did, they would emit much more ultra high energy cosmic rays than we observe. If $T_{eq}\ll 13\% \times 10^{15}$GeV, then it is still possible that ultra light black holes form the majority of dark matter however, they cannot be the source of the ultra high energy cosmic rays which we observe because they will not radiate enough. Only if $T_{eq}\approx 13\% \times 10^{15}$GeV can we have that dark matter consist mostly of ultra light black holes and that those black holes are simultaneously the source for the observed ultra high energy cosmic rays. Interestingly, it turns out that $T_{eq}\approx 13\% \times 10^{15}$GeV is consistent with theory \cite{mukhanov}.
\section*{Conclusions \& Discussion}
In this paper we have studied the new
Reissner-Nordstr\"om-like metric obtained in the paper \cite{RNR}
We recall the LQBH metric
%\begin{widetext}
%\begin{centering}
\begin{eqnarray}
%&& \hspace{-0.5cm}
&& \hspace{-0.3cm} ds^2 = -\frac{ (r - r_+) (r-r_-)(r+ r_{\star} )^2 }{ r^4 + a_0^2}dt^2
\nonumber \\
&& \hspace{0.2cm}
+ \frac{dr^2}{\frac{(r-r_+)(r-r_-)r^4}{(r+ r_{\star}))^2 (r^4 + a_0^2)}}
+ \Big(\frac{a_0^2}{ r^2} + r^2\Big) d  \Omega^{(2)}.
%\nonumber \\
%&&
\label{metricabella2}
\end{eqnarray}
%\end{centering}
%\end{widetext}
%\end{centering}
%Relative to the black hole singularity problem the main
%results are
%The LQBH's metric (\ref{metricabella2})  properties has the following
%\begin{enumerate}
%\item $ \lim_{r \rightarrow +\infty} g_{\mu \nu}(r) = \eta_{\mu \nu}$,
%\item  $ \lim_{r \rightarrow 0} g_{\mu \nu}(r) = \eta_{\mu \nu}$,
%\item $\lim_{m, a_0 \rightarrow 0} g_{\mu \nu}(r)  = \eta_{\mu \nu}$,
%\item $K(g) < \infty \,\, \forall r$,
%\item $r_{\rm Max}(K(g)) \sim \sqrt{a_0}$.
%\end{enumerate}
%In particular (see point 5.) {\em the position $(r_{\rm Max})$
%where the Kretschmann  invariant  operator %$K(g)$
%is maximum is independent from the black hole mass and from
%the polymeric parameter $\delta_b$}.
The metric has two event horizons that we have defined $r_+$ and $r_-$;
$r_+$ is the Schwarzschild event horizon and $r_-$ is an inside horizon
tuned by the polymeric parameter $\delta_b$.
The solution has many similarities with the Reissner-Nordstr\"om metric
but without curvature singularities anywhere. In particular the region $r=0$ corresponds to
another asymptotically flat region. No massive particle can reach this region in a finite proper time. A careful analysis shows that
the metric has a {\em Schwarzschild core} in $r\approx 0$  of mass $M\propto a_0/m$.
%Our analysis shows that the %What solve the
%The crucial point in the
%singularity problem is solved by  %resolution is %to have
%a bounce of the $S^2$ sphere on a minimum area $a_0 >0$.
We have studied the black hole thermodynamics : temperature, entropy
%In this paper we have concentrated our attention on the
and the evaporation process.
%and we have calculated the temperature, entropy ( with all the correction suggested by
%the particular model) and the mass variation formula as seen by a distant observer at
%the time $v$.
The main results are the following.
%\begin{enumerate}
%= \frac{48 M^2 G_N^2}{b(t)^6} \hspace{0.5cm} \rightarrow
%\hspace{0.5cm} \widehat{\mathcal{R}_{\mu \nu \rho \sigma} \,
%\mathcal{R}^{\mu \nu \rho \sigma}} |0 \rangle = \widehat{\frac{48 M^2
%G_N^2}{|x|^6}} |0 \rangle = \frac{384 M^2 G_N^2}{\pi^3 l_P^6} |0
%\rangle. \nonumber
%\end{eqnarray}
%\item
The temperature $T(m)$ goes to zero
for $m \approx 0$ and reduces to the Bekenstein-Hawking temperature for large
values of the mass
Bekenstein-Hawking
%\begin{eqnarray}
$T(m) = 128 \pi  \, m^3/[1024 \pi^2 m^4 + A_{Min}^2]$.
%\label{TempDis}
%\end{eqnarray}
The black hole entropy in terms of the event horizon area
%and the LQG minimum area eigenvalue is
%reproduces the $A/4l_P^2$ term but contain also the $\ln$-correction and
%all the other correction in $(l_P^2/A)^n$
%\begin{eqnarray}
is $S = \sqrt{A^2 - A_{Min}^2}/4$.
% \label{SC}
 %\end{eqnarray}
%(where we have repristed the length units).
The evaporation process needs an infinite time in our semiclassical analysis
but the difference with the classical result is evident only at the Planck scale. The fact that the black holes can never fully evaporate resolves the information loss paradox. Furthermore, we showed that because of the temperature profile of the LQBH, the fact that the temperature decreases for very light black holes, a black hole in thermal environment will never totally evaporate but will thermalise with the background. The CMB is such a background that can stabilise the ultra-light black holes. Since the horizon of ultra-light LQBH is hidden behind a wormhole with Planck size cross section, cold and light black holes could act as very weakly interacting dark matter. However the universe is not old enough for black holes created during the Big Bang to have cooled down to $2.7$ K, they would still be excessively hot.

We know that in the very early universe ultra-light black holes cannot be created because the universe expands at rate which is much faster than the rate of collisions between particle. Particles of the Standard Model start colliding together at a rate faster than expansion of the universe when the temperature has cool lower than $10^{15}$GeV$-10^{17}$GeV. If we suppose that the temperature $T_{eq}$ at which local equilibrium of the matter is achieved and thus black holes can be formed from fluctuations of the matter is $13\%$ of $10^{15}$GeV then ultra-light black holes can explain both dark matter and cosmic rays with energies above the GZK cut off.

We would have that once the universe has cooled to $1.3\times10^{14}$GeV, ultra-light black holes, the overwhelming majority of which have a mass inferior to $5\times 10^{-5} m_P$ would be created from fluctuations of the matter. These black holes are still very hot and radiate, but because they are hidden behind a Planck-sized wormhole, they do so very slowly and on average would lose less than 1 part in $10^{14}$ of there mass since their creation and are for all practical purposes stable.   If since their creation the universe has expanded by a scale factor of $10^{37}$ the mass of all these
ultra-light black holes would exactly equal the mass of dark matter and they could explain the entirety of dark matter.

Since the universe has expanded by a scale factor of about $10^{28}$ since the end of inflation, and that it expanded by a scale factor of at least $10^{28}$ during inflation, the fact that universe has expanded by a scale factor of $10^{37}$ since the birth of the black holes would mean that the black holes would have been created during inflation. This in turn would mean that inflation would  still be underway when the universe had temperature of $1.3\times 10^{14}$Gev. %This is not far form
This is very close to
the simplest models of inflation which situate inflation at energy scales of $10^{15}$GeV$-10^{16}$GeV.

In turn, if the black whole were created when the universe was at a temperature of $1.3\times 10^{14}$GeV, then the amount of cosmic rays with energies higher than the GZK cut off they would emit would correspond exactly to the amount of such radiation observed. Because they interact with the CMB, cosmic rays cannot travel more than 50 Mpc before seeing their energy fall below the GZK cut off: $6\times 10^{19}$eV. However we do see particles with energies above the GZK cut off but we do not see any sources for such energetic particles within 50 Mpc from us. These energetic particles, dubbed ultra high energy cosmic rays are thus a mystery for the moment.

{\em Hence in conclusion}, ultra-light LQG black holes have the potential to resolve two outstanding problems in physics: what is dark matter, and where do ultra high energy cosmic rays come from. It is also noteworthy that much of these results do not actually depend on exact details of the black holes. The essential feature is that the temperature of the black holes goes to zero when their mass goes to zero, the results being very generic. It is thus likely that the same effect could be observed with non-commutative black holes and asymptotic safety gravity black holes \cite{tempmod} \cite{BR}, both of which exhibit zero temperature at zero mass or for a remnant mass.
The same analysis we think could be applied to the new Ho\v{r}ava-Lifshitz quantum gravity \cite{HL}.

\section*{Acknowledgements}

%We are strongly indebted to Roberto Balbinot
%for crucial criticisms, inputs and suggestions.
We are extremely grateful to the fantastic environment offered by Perimeter Institute.
We are grateful to Johannes Knapp explaining to the authors the inner workings of the Auger
detector. %the help given in the
%for many important and clarifying discussions.
The authors would also like to thank Niayesh Afshordi and Neil Turok for kindly answering their cosmology questions.
%We are extremely grateful to the fantastic environment offered by Perimeter Institute.
Research at
Perimeter Institute is supported by the Government of Canada through Industry Canada
and by the Province of Ontario through the Ministry of Research \& Innovation.


\begin{thebibliography}{99}
\bibitem{book}
C. Rovelli, {\em Quantum Gravity}, (Cambridge University Press,
Cambridge, 2004);
A. Ashtekar, {\em Background independent quantum gravity: A Status report},
Class. Quant. Grav. 21, R53 (2004), gr-qc/0404018;
T. Thiemann, {\em Loop quantum gravity: an inside view}, hep-th/0608210;
T. Thiemann, {\em Introduction to Modern Canonical Quantum General Relativity},
gr -qc/0110034; {\em Lectures on Loop Quantum Gravity},
Lect. Notes Phys. 631, 41-135 (2003), gr-qc/0210094.
%
\bibitem{SpinFoams} F. Conrady, L. Freidel,
{\em Quantum geometry from phase space reduction},
[arXiv:0902.0351];
J. W. Barrett, R. J. Dowdall, W. J. Fairbairn, H. Gomes, F. Hellmann,
{\em Asymptotic analysis of the EPRL four-simplex amplitude},
[arXiv:0902.1170];
F. Conrady, L. Freidel,
{\em Path integral representation of spin foam models of 4d gravity},
Class. Quant. Grav.25, 245010, 2008,
[arXiv:0806.4640];
F. Conrady, L. Freidel,
{\em On the semiclassical limit of 4d spin foam models},
Phys. Rev. D78, 104023, 2008,
[arXiv:0809.2280];
J. Engle, E. Livine, R. Pereira, C. Rovelli,
{\em LQG vertex with finite Immirzi parameter},
Nucl. Phys. B799 (2008) 136-149,
[arXiv:0711.0146];
J. Engle, R. Pereira, C. Rovelli,
{\em Flipped spinfoam vertex and loop gravity}
Nucl. Phys. B798 (2008) 251-290,
[arXiv:0708.1236];
J. Engle, R. Pereira, C. Rovelli,
{\em The Loop-quantum-gravity vertex-amplitude},
Phys. Rev. Lett. 99 (2007) 161301,
[arXiv:0705.2388];
%\bibitem{SemiLim}
E. Alesci, C. Rovelli,
{\em The Complete LQG propagator I. Difficulties with the Barrett-Crane vertex},
Phys. Rev. D76 (2007) 104012,
[arXiv:0708.0883];
S. Speziale,
{\em Background-free propagation in loop quantum gravity}
[arXiv:0810.1978];
E. Bianchi, L. Modesto, C. Rovelli, S. Speziale
{\em Graviton propagator in loop quantum gravity},
Class. Quant. Grav. 23 (2006) 6989-7028,
[gr-qc/0604044];
L. Modesto, C. Rovelli
{\em Particle scattering in loop quantum gravity}
in Phys. Rev. Lett. 95 (2005) 191301,
[gr-qc/0502036];
E. Bianchi, L. Modesto,
{\em The Perturbative Regge-calculus regime of loop quantum gravity},
Nucl. Phys. B 796 (2008) 581-621,
[arXiv:0709.2051];
E. Bianchi, A. Satz,
{\em Semiclassical regime of Regge calculus and spin foams},
[arXiv:0808.1107].
\bibitem{Fracton} L. Modesto, {\em Fractal Structure of Loop Quantum Gravity},
[arXiv:0812.2214];
L. Modesto,
{\em Fractal Quantum Space-Time},
[arXiv:0905.1665];
F. Caravelli, L. Modesto,
{\em Fractal Dimension in 3d Spin-Foams},
[arXiv:0905.2170].
%\bibitem{MR} Martin Reuter, {\em Non perturbative
%evolution equation for quantum gravity}, Phys. Rev. D57 971-985 (1998), hep-th/9605030
%\bibitem{MV} F. Conrady, L. Doplicher, R. Oeckl, C. Rovelli and M. Testa
%{\em Minkowski vacuum in background independent quantum gravity}, Phys. Rev. D 69 064019,
%gr-qc/0307118;
%Daniele Colosi, Luisa Doplicher, Winston Fairbairn, Leonardo Modesto,
%Karim Noui and Carlo Rovelli, {\em Background independence in a nutshell: dynamics of a tetrahedron}, Class. Quant.
%Grav. 22 (2005) 2971-2989, gr-qc/0408079
%bibitem{ModestoRovelli} Leonardo Modesto and Carlo Rovelli, {\em Particle scattering
%in loop quantum gravity}, Phys. Rev. Lett. 95 191301 (2005), gr-qc/0502036
%\bibitem{Rovelli1} Carlo Rovelli, {\em Graviton propagator from background-independent
%quantum gravity},Phys. Rev. Lett. 97 151301 (2006), gr-qc/0508124
% \bibitem{BMSR} Eugenio Bianchi, Leonardo Modesto, Carlo Rovelli and Simone Speziale,
% {\em Graviton propagator in loop quantum gravity}, Class. Quant. Grav. 23 (2006) 6989 -7028,
% gr-qc/0604044
 %\bibitem{Simone1} Simone Speziale, {\em Towards the graviton from spinfoams :
 %the 3D toy model}, J. high Energy Phys. JHEP05 (2006) 039, gr-qc/0512102
%\bibitem{Simone2} E. Livine and S. Speziale, {\em Group integral techniques for the spinfoam graviton propagator},
%gr-qc/0608131
%\bibitem{Freidel} Laurent Freidel and Etera R. Livine  {\em Ponzano-Regge model revisited III: Feynman diagrams and
%effective field theory}, Class. Quant. Grav. 23 2021-2062 (2006), hep-th/0502106
  %\bibitem{FB} Aristide Baratin and Laurent Freidel, {\em Hidden quantum gravity in 4d Feynman diagrams: emergence of
  %spin foams}, hep-th/0611042
 %Aristide Baratin and Lourent Freidel, {\em Hidden quantum gravity in 4d Feynman diagrams :
%Emergence of spin foams}, gr-qc/0611042
%\bibitem{Winston} Winston J. Fairbairn {\em Fermions in three-dimensional spinfoam quantum gravity}, gr-qc/0609040
%\bibitem{FreidelAristide}
\bibitem{Boj}
M. Bojowald,  {\em Loop quantum cosmology},
Living Rev. Rel. 8:11, 2005,
[gr-qc/0601085];
M. Bojowald,
{\rm Absence of singularity in loop quantum cosmology}
Phys. Rev. Lett. 86:5227-5230, 2001,
[gr-qc/0102069].
\bibitem{MAT} A. Ashtekar, M. Bojowald and J. Lewandowski, {\em Mathematica structure of loop quantum cosmology}, Adv.
    Theor. Math. Phys. 7 (2003) 233-268, [gr-qc/0304074].
%A. Ashtekar , Tomasz Pawlowski, Parampreet Singh, Kevin Vandersloot,
%{\em Loop quantum cosmology of k=1 FRW models},
%Phys. Rev. D75 (2007) 024035, gr-qc/0612104 %;
%A. Ashtekar, T. Pawlowski, Parampreet Singh,
%{\em Quantum Nature of the Big Bang: Improved dynamics},
%Phys. Rev. D74 (2006) 084003,
%gr-qc/0607039;
%Abhay Ashtekar, Tomasz Pawlowski, Parampreet Singh,
%{\em Quantum Nature of the Big Bang: An Analytical and Numerical Investigation. I},
%Phys.Rev.D73 (2006) 124038,
%gr-qc/0604013
\bibitem{KS} R. Kantowski and R. K. Sachs, J. Math. Phys. 7 (3) (1966);
L. Bombelli \& R. J. Torrence {\em Perfect fluids and Ashtekar variables,
with application to Kantowski-Sachs models},
 Class. Quant. Grav. 7 (1990) 1747-1745.
 \bibitem{h} K. V. Kuchar,
{\em Geometrodynamics of Schwarzschild Black Hole},
Phys. Rev. D50 (19994) 3961-3981,
[gr-qc/9403003];
T. Thiemann,
{\em Reduced models for quantum gravity},
Lect. Notes Phys. 434 (1994) 289-318,
[gr-qc/9910010];
H.A. Kastrup, T. Thiemann,
{\em Spherically symmetric gravity as a completely integrable system},
Nucl. Phys. B 425 (1994) 665-686,
[gr-qc/9401032];
T. Thiemann, H.A. Kastrup,
{\em Canonical quantization of spherically symmetric gravity in Ashtekar's selfdual representation}
Nucl. Phys. B 399 (1993) 211-258,
[gr-qc/9310012].
\bibitem{work1} L. Modesto, {\em Disappearance of the black hole singularity in loop  quantum gravity}, Phys.
    Rev. D 70 (2004) 124009, [gr-qc/0407097];
L. Modesto, {\em The kantowski-Sachs space-time in loop quantum gravity}, International
Int. J. Theor. Phys. 45 (2006) 2235-2246,
 [gr-qc/0411032];
 L. Modesto,
{\em  Loop quantum gravity and black hole singularity},
proceedings of 17th SIGRAV Conference, Turin, Italy, 4-7 Sep 2006,
[hep-th/0701239];
L. Modesto, {\em Gravitational collapse in loop quantum gravity},
Int. J. Theor. Phys.47 (2008) 357-373,
[gr-qc/0610074];
L. Modesto, {\em Quantum gravitational collapse}, [gr-qc/0504043].
\bibitem{work2}
A. Ashtekar and M. Bojowald,
{\em Quantum geometry and Schwarzschild singularity} Class. Quant. Grav. 23 (2006) 391-411,
[gr-qc/0509075];  L. Modesto, {\em Loop quantum black hole}, Class. Quant. Grav. 23 (2006) 5587-5602,
[gr-qc/0509078].
\bibitem{GP}
R. Gambini, J. Pullin,
{\rm Black holes in loop quantum gravity: The Complete space-time},
Phys. Rev. Lett.101:161301, 2008,
[arXiv:0805.1187];
M. Campiglia, R. Gambini, J. Pullin,
{\em Loop quantization of spherically symmetric midi-superspaces : the interior problem},
AIP Conf. Proc. 977:52-63, 2008,
[arXiv:0712.0817];
M. Campiglia, R. Gambini, J. Pullin,
Loop quantization of spherically symmetric midi-superspaces
Class. Quant. Grav. 24:3649-3672, 2007,
[gr-qc/0703135].
\bibitem{Sabine}
  S. Hossenfelder and L. Smolin,
  ``Conservative solutions to the black hole information problem,''
  [arXiv:0901.3156];
  A. Ashtekar, V. Taveras, M. Varadarajan,
  {\em Information is Not Lost in the Evaporation of 2-dimensional Black Holes},
  Phys. Rev. Lett. 100, 211302, 2008,
  [arXiv:0801.1811].
  %%CITATION = ARXIV:0901.3156;%%
\bibitem{SS} L. Modesto, {\em Black hole interior from loop quantum gravity},
Advances in High Energy Physics Volume 2008 (2008), Article ID 459290
[gr-qc/06011043]; L. Modesto,
{\em Evaporating loop quantum black hole},
[gr-qc/0612084].
\bibitem{SS2}
C. G. Bohmer, K.Vandersloot,
{\em Loop quantum dynamics of the Schwarzschild Interior},
[arXiv:0709.2129];
%\bibitem{SS3}
Dah-Wei Chiou,
{\em Phenomenological Loop Quantum Geometry of the Schwarzschild Black Hole},
arXiv:0807.0665.
\bibitem{LoopOld}  C. Rovelli and L. Smolin,
  {\em Loop Space Representation Of Quantum General Relativity},
  Nucl. Phys. B 331 (1990) 80;
  %%CITATION = NUPHA,B331,80;%%
   C. Rovelli and L. Smolin,
  {\em Discreteness of area and volume in quantum gravity},
  Nucl.  Phys. B 442 (1995) 593;
  E. Bianchi,
 {\em  The Length operator in Loop Quantum Gravity},
 Nucl. Phys. B 807 (2009) 591-624,
[arXiv:0806.4710].
\bibitem{RNR} L. Modesto,
{\em Space-Time Structure of Loop Quantum Black Hole},
arXiv:0811.2196.
%\cite{Hossenfelder:2009xq}
\bibitem{variables} A. Ashtekar, {\em New Hamiltonian formulation of general relativity},
Phys. Rev. D 36 1587-1602.
%
\bibitem{HE} S.W. Hawking, G.F.R. Ellis
{\em The large scale structure of space-time}
Cambridge University Press, 1973.
\bibitem{BR}
A. Bonanno, M. Reuter, {\em Renormalization group improved black hole space-times}, Phys. Rev. D
    62
    (2000) 043008, [hep-th/0002196];
A. Bonanno, M. Reuter
{\em Spacetime structure of an evaporating black hole in quantum gravity},
Phys. Rev. D 73 (2006) 083005,
[hep-th/0602159].
\bibitem{AM} M. Arzano,
{\em Black hole entropy, log corrections and quantum ergosphere}
Phys.Lett.B634 (2006) 536-540,
[gr-qc/0512071];
Mi. Arzano, A.J.M. Medved, E. C. Vagenas,
{\em Hawking radiation as tunneling through the quantum horizon}
JHEP 0509 (2005) 037,
[hep-th/0505266];
R. Banerjee, B. Ranjan Majhi,
{\em Quantum Tunneling and Back Reaction},
Phys. Lett. B 662 (2008) 62-65,
[arXiv:0801.0200];
R. Banerjee, B. Ranjan Majhi
{\em Quantum Tunneling Beyond Semiclassical Approximation}
JHEP 0806 (2008) 095,
[arXiv:0805.2220]
R. Banerjee, B. Ranjan Majhi,
{\em Quantum Tunneling and Trace Anomaly}
[arXiv:0808.3688];
B. Ranjan Majhi,
{\em Fermion Tunneling Beyond Semiclassical Approximation},
[arXiv:0809.1508].
\bibitem{Fabbri} A. Fabbri and J. Navarro-Salas,
{\em Modeling Black Hole Evaporation},
Imperial College Press (2005).
 \bibitem{AB} A. Ashtekar \& M. Bojowald, {\em Black hole evaporation : A paradigm}
Class. Quant. Grav. 22 (2005) 3349-3362, [gr-qc/0504029].
\bibitem{PInsta} M. Simpson and R. Penrose,
{\em Internal instability in Reissner-Nordstr\"om black hole},
Int. J. Theor. Phys. 7 (1973) 183-197.
\bibitem{DO} D. Oriti,
{\em Group field theory as the microscopic description of the quantum spacetime fluid: A New perspective on the
continuum in quantum gravity}
[arXiv:0710.3276].
%\bibitem{Bojk=1}
%\bibitem{Bombelli}
%L. Bombelli \& R. J. Torrence,  ``Perfect fluids and Ashtekar variables,
%with application to Kantowski-Sachs models", Class. Quant. Grav. {\bf 7} (1990) 1747-1745
%\bibitem{BojRP} Martin Bojowald,  ``Loop quantum cosmology: recent progress", gr-qc/0402053
%\bibitem{I.R} Ioannis Raptis, {\em Finitary-Algebraic `Resolution' of the Inner Schwarzschild
%Singularity}, gr-qc/0408045; Anastasios Mallios and Ioannis Raptis,{\em Smooth Singularities Exposed: Chimeras of the Differential Spacetime   Manifold}, gr-qc/041112;  Ioannis Raptis,
%{\em `Iconoclastic', Categorical Quantum Gravity}, gr-qc/0509089
\bibitem{Bombelli} L. Bombelli \& R. J. Torrence {\em Perfect fluids and Ashtekar variables,
with application to Kantowski-Sachs models} Class. Quant. Grav. {\bf 7} (1990) 1747-1745.
%\bibitem{AFW} A. Ashtekar, S. Fairhurst and J. Willis, {\em Quantum gravity, shadow states, and quantum mechanics},
%Class. Quant. Grav. {\bf 20} 1031-1062 (2003)
%bibitem{Partial} C Rovelli, ``Partial observables", {Phy
%Rev} {D65} (2002) 124013; gr-qc/0110035
%\bibitem{Fonte} Viqar Husain and Oliver Winkler, {\em On singularity resolution in quantum gravity} ; gr-qc/0312094
%\bibitem{Fonte.Math} H. Halvorson, {\em Complementary of representations in quantum mechanics}, Studies in History and Philosophy of Modern Physics 35, 45-56   (2004), quant - ph/0110102
%\bibitem{Thie} T. Thiemann, {\em Quantum Spin Dynamics}, Class. Quant. Grav. {\bf 15}, 839 (1998)
%\bibitem{Thie.2} T. Thiemann, {\em Introduction to Modern Canonical Quantum General Relativity} ; gr -qc/0110034; {\em Lectures on Loop Quantum Gravity}, gr-qc/0210094
%\bibitem{BojInverse} Martin Bojowald, {\em Quantization ambiguities in isotropic
%quantum geometry}; gr-qc/0206053
%\bibitem{Smolin}  Sundance O. Bilson-Thompson,  Fotini Markopoulou and Lee Smolin,
%{\em Quantum gravity and the standard model}, hep-th/0603022
%\cite{Kapusta:1984yh}
\bibitem{Kapu}
  J. I. Kapusta,
  {\em Nucleation Rate for Black Holes}
 %``NUCLEATION RATE FOR BLACK HOLES,''
  Phys. Rev. D {\bf 30} (1984) 831.
  %%CITATION = PHRVA,D30,831;%%
%\cite{Zhang:2007bi}
\bibitem{yiling}
  X. Zhang and Y. Ling,
  ``Inflationary universe in loop quantum cosmology,''
  JCAP  0708 (2007) 012
  [gr-qc/0705.2656]. % [gr-qc]].
  %%CITATION = JCAPA,0708,012;%%
%\cite{Meissner:2004ju}
\bibitem{mukhanov}
V.~Mukhanov, \emph{Physical Foundations of Cosmology,}(Cambridge University Press,
Cambridge, 2005), pp. 74-75.
\bibitem{gamma}
  K. A. Meissner,
  ``Black hole entropy in loop quantum gravity,''
  Class. Quant. Grav.  21 (2004) 5245
  [gr-qc/0407052].
  %%CITATION = CQGRD,21,5245;%%
%\cite{Greisen:1966jv}
\bibitem{GZK}
  K. Greisen,
  {\em End To The Cosmic Ray Spectrum?},
  Phys.  Rev. Lett.  16, 748 (1966);
  %%CITATION = PRLTA,16,748;%%
  G. T. Zatsepin and V. A. Kuzmin,
  {\em Upper limit of the spectrum of cosmic rays},
  JETP Lett.  4, 78 (1966)
  [Pisma Zh. Eksp.  Teor. Fiz. 4, 114 (1966)].
  %%CITATION = ZFPRA,4,114;%%
%\cite{Berezinsky:1997hy}
\bibitem{HiRes}
  R. Abbasi {\it et al.}  [HiRes Collaboration],
  {\em Observation of the GZK cutoff by the HiRes experiment},
  Phys. Rev. Lett. 100 (2008) 101101
  [astro-ph/0703099];
    J.~Abraham {\it et al.}  [Pierre Auger Collaboration],
  {\em Observation of the suppression of the flux of cosmic rays above $4\times 10^{19}$eV},
  Phys.\ Rev.\ Lett.\  {\bf 101} (2008) 061101
  [arXiv:0806.4302].
  %%CITATION = PRLTA,79,4302;%%
%\cite{Bourjaily:2004aj}
\bibitem{jacob}
  J. L. Bourjaily,
  {\em Determining the actual local density of dark matter particles},
  Eur.  Phys. J.  C 40N6 (2005) 23
  [astro-ph/0410470].
  %%CITATION = EPHJA,C40N6,23;%%
  %% QUI
  \bibitem{Xpart}
  V. Berezinsky, M. Kachelriess and A. Vilenkin,
  {\em Ultra-high energy cosmic rays without GZK cutoff},
  Phys. Rev. Lett. 79 (1997) 4302
  [astro-ph/9708217].
%\bibitem{HiRes}
%  R. Abbasi {\it et al.}  [HiRes Collaboration],
%  {\em Observation of the GZK cutoff by the HiRes experiment},
  %Phys. Rev. Lett. 100 (2008) 101101
 % [arXiv:astro-ph/0703099];
  %\cite{Abraham:2008ru}
%
%\bibitem{Abraham:2008ru}
%  J.~Abraham {\it et al.}  [Pierre Auger Collaboration],
%  ``Observation of the suppression of the flux of cosmic rays above $4\times 10^{19}$eV,''
 % Phys.\ Rev.\ Lett.\  {\bf 101} (2008) 061101
%  [arXiv:0806.4302 [astro-ph]].
  %%CITATION = PRLTA,101,061101;%%
\bibitem{tempmod}
Yun Soo Myung, Yong-Wan Kim, Young-Jai Park,
{\em Thermodynamics of regular black hole},
[arXiv:0708.3145];
Yun Soo Myung, Yong-Wan Kim, Young-Jai Park,
{\em Quantum Cooling Evaporation Process in Regular Black Holes}
Phys. Lett. B (2007) 656:221-225,
[gr-qc/0702145];
P. Nicolini,
{\em Noncommutative Black Holes, The Final Appeal To Quantum Gravity: A Review},
[arXiv:0807.1939];
R. Banerjee, B. Ranjan Majhi, S. Samanta,
{\em Noncommutative Black Hole Thermodynamics},
Phys. Rev. D77 (2008) 124035,
[arXiv:0801.3583];
Banerjee, Bibhas Ranjan Majhi, Sujoy Kumar Modak,
{\em Area Law in Noncommutative Schwarzschild Black Hole},
[arXiv:0802.2176].
\bibitem{HL} P. Ho\v{r}ava,
{\em Quantum Gravity at a Lifshitz Point},
Phys. Rev. D 79, 084008, 2009,
[arXiv:0901.3775];
Gianluca Calcagni, {\em Cosmology of the Lifshitz universe},
[arXiv:0904.0829];
Y. S. Myung, Y. W. Kim,
{\em Thermodynamics of Ho\v{r}ava-Lifshitz black holes}
[arXiv:0905.0179];
R. G. Cai, L. M. Cao, N. Ohta,
{\em Thermodynamics of Black Holes in Ho\v{r}ava-Lifshitz Gravity},
[arXiv:0905.0751];
Y. S. Myung,
{\em Thermodynamics of black holes in the deformed Ho\v{r}ava-Lifshitz gravity},
[arXiv:0905.0957].
%Non-cummutative blackholes and martin Reuter black holes.
  %%CITATION = PRLTA,100,101101;%%
  %
  %
%\bibitem{Rosquist}
 % K. Rosquist,
%{\em Gravitationally induced electromagnetism at the Compton scale},
%  Class. Quant. Grav.  23 (2006) 3111
%  [arXiv:gr-qc/0412064].
  %%CITATION = CQGRD,23,3111;%%







\end{thebibliography}
\end{document}